\documentclass[aps,prc,onecolumn,groupaddress,showpacs,nofootinbib,floatfix]{revtex4}

\usepackage{bm}
\usepackage{epsfig}
\usepackage{longtable}

\def\bea {\begin{eqnarray}}
\def\eea {\end{eqnarray}}
\def\be {\begin{equation}}
\def\ee {\end{equation}}
\def\ben{\begin{enumerate}}
\def\een{\end{enumerate}}
\def\bi{\begin{itemize}}
\def\ei{\end{itemize}}

\def\eg{{\it e.g.}\ }
\def\etal{{\it et al.}\ }

\def\F{{\cal F}}
\def\prl {Phys. Rev. Lett.\ }
\def\pl {Phys. Lett.\ }
\def\pr {Phys. Rev.\ }
\def\np {Nucl. Phys.\ }
\def\gV{g_{\mbox{\tiny V}}}

\def\fM{f_{\mbox{\tiny M}}}

\def\fS{f_{\mbox{\tiny S}}}

\def\GV{G_{\mbox{\tiny V}}}
\def\GF{G_{\mbox{\tiny F}}}
\def\DRV{\Delta_{\mbox{\tiny R}}^{\mbox{\tiny V}}}

\newcommand{\sfrac}[2]{\mbox{\small{$\frac{#1}{#2}$}}}

\begin{document} 

\title{Superallowed $0^+$$\rightarrow 0^+$ nuclear $\beta$ decays:
A new survey with precision tests of the conserved vector current hypothesis and the standard model}

\author{J.C. Hardy} 
\email{hardy@comp.tamu.edu}
\author{I.S. Towner}
\affiliation{Cyclotron Institute, Texas A\&M University, College Station, Texas 77843}

\date{\today}

\begin{abstract}
A new critical survey is presented of all half-life, decay-energy and branching-ratio
measurements related to 20 superallowed $0^+$$\rightarrow 0^+$ $\beta$ decays.  Compared with our
last review, there are numerous improvements: First, we have added 27 recently published
measurements and eliminated 9 references, either because they have been superceded by much
more precise modern results or because there are now reasons to consider them fatally
flawed; of particular importance, the new data include a number of high-precision Penning-trap
measurements of decay energies.  Second, we have used the recently improved isospin
symmetry-breaking corrections, which were motivated by these new Penning-trap results.  Third,
our calculation of the statistical rate function $f$ now accounts for possible excitation
in the daughter atom, a small effect but one which merits inclusion at the present level of
experimental precision.  Finally, we have re-examined the systematic uncertainty associated
with the isospin symmetry-breaking corrections by evaluating the radial-overlap correction
using Hartree-Fock radial wave functions and comparing the results with our earlier
calculations, which used Saxon-Woods wave functions; the provision for systematic uncertainty
has been changed as a consequence.  The new ``corrected" $\F t$ values are impressively
constant and their average, when combined with the muon liftime, yields the up-down
quark-mixing element of the Cabibbo-Kobayashi-Maskawa (CKM) matrix, $V_{ud} = 0.97425 \pm 0.00022$.
The unitarity test on the top row of the matrix becomes
$|V_{ud}|^2 + |V_{us}|^2 + |V_{ub}|^2 = 0.99995 \pm 0.00061$.  Both $V_{ud}$ and the unitarity sum
have significantly reduced uncertainties compared with our previous survey, although the new value
of $V_{ud}$ is statistically consistent with the old one.  From these data we also set limits on the
possible existence of scalar interactions, right-hand currents and extra $Z$ bosons.  Finally, we
discuss the priorities for future theoretical and experimental work with the goal of making the CKM
unitarity test even more definitive.

\end{abstract}

\pacs{23.40.Bw, 12.15.Hh, 12.60.-i}
\maketitle

\section{\label{s:intro} Introduction}

Precise measurements of the beta decay between nuclear analog states of spin, $J^{\pi} = 0^+$,
and isospin, $T = 1$, provide demanding and fundamental tests of the properties of the
electroweak interaction.  Collectively, these transitions can sensitively probe the conservation
of the vector weak current, set tight limits on the presence of scalar or right-hand currents
and, by providing the most precise value for $V_{ud}$, the up-down quark-mixing element of
the Cabibbo-Kobayashi-Maskawa (CKM) matrix, they can contribute to the most demanding available
test of the unitarity of that matrix, a property which is fundamental to the electroweak standard model.

We have published five previous surveys of $0^+$$\rightarrow 0^+$ superallowed transitions
\cite{TH73,HT75,Ko84,HT90,HT05}, the first having appeared 35 years ago and the most recent, four
years ago.  In each, we published a complete survey of all relevant nuclear data that pertained to these
superallowed transitions and used the results to set limits on the weak-interaction parameters that
were important at the time.  A particularly noteable outcome of our analysis four years ago \cite{HT05}
was that the sum of squares of the top-row elements of the CKM matrix -- the test of CKM unitarity --
remained ambiguous, with the possibility of a significant shortfall in the unitarity sum.

Since our last survey closed in November 2004, there has been a great deal of activity in this
field prompted at least in part by the tantalizing possibility that new physics could be revealed
by a failure in CKM unitarity.  New measurements relating to $0^+$$\rightarrow 0^+$ superallowed
transitions have appeared in 27 publications, an addition of 20\% to the papers accumulated up to 2004.
Many of these measurements were of unprecedented precision so they did not merely add more of the
same: they palpably improved the results, in some cases by tightening their error bars and, in others,
by changing their central values.  Penning-trap measurements of decay energies, which only
became possible after 2004, have been especially effective in this regard.

In addition to new measurements, there have also been important improvements to the small theoretical
corrections that must be applied to the data in order to extract $V_{ud}$ and the values of other
weak-interaction parameters.  In the past four years, the radiative \cite{Ma06} and isospin
symmetry-breaking corrections \cite{TH08} have both been subjected to major re-evaluations, which
have undoubtedly improved their values and, in the former case, has reduced the uncertainty by a
factor of two.

In parallel with these developments, there has also been considerable activity in the determination
of $V_{us}$, the other matrix element that plays a role in the top-row unitarity test of the CKM matrix.
(The third element in the top row, $V_{ub}$, is very small and contributes a negligible 0.001\% to
the unitarity sum.) As with the work related to $V_{ud}$, this activity has encompassed new
experiments -- precise measurements of kaon branching ratios -- as well as improved theoretical corrections.
However, in contrast with $V_{ud}$, not only the uncertainty of $V_{us}$ but also its central value
have been considerably changed by this recent work (see \cite{Fl08} for an up-to-date overview of
$V_{us}$).

Overall, the recent improvements have been numerous enough and their impact on the unitarity test
significant enough that this is an opportune time to produce a new and updated survey of the nuclear
data used to establish $V_{ud}$.  We incorporate data on a total of 20 superallowed transitions and
have continued the practice we began in 1984 \cite{Ko84} of updating all original data to take account
of the most modern calibration standards.  In addition to including the improved correction terms already
referred to, we have also upgraded our calculation of the statistical rate function $f$ to include
provision for excitation of the daughter atom, and we have included a more extensive treatment of
possible systematic uncertainties associated with the isospin symmetry-breaking corrections.

Superallowed $0^+$$\rightarrow 0^+$ $\beta$ decay between $T=1$ analog states depends uniquely on
the vector part of the weak interaction and, according to the conserved vector current (CVC) hypothesis,
its experimental $ft$ value should be related to the vector coupling constant, a fundamental constant
which is the same for all such transitions.  In practice, the expression for $ft$ includes several
small ($\sim$1\%) correction terms.  It is convenient to combine some of these terms with the
$ft$ value and define a ``corrected" $\F t$ value.  Thus, we write \cite{HT05}
\be
\F t \equiv ft (1 + \delta_R^{\prime}) (1 + \delta_{NS} - \delta_C ) = \frac{K}{2 \GV^2 
(1 + \DRV )}~,
\label{Ftconst}
\ee
where $K/(\hbar c )^6 = 2 \pi^3 \hbar \ln 2 / (m_e c^2)^5 = 8120.2787(11) \times
10^{-10}$ GeV$^{-4}$s, $\GV $ is the vector coupling constant for semi-leptonic weak interactions,
$\delta_C$ is the isospin-symmetry-breaking correction and $\DRV$ is the transition-independent part
of the radiative correction.  The terms $\delta_R^{\prime}$ and $\delta_{NS}$ comprise the
transition-dependent part of the radiative correction, the former being a function only of the
electron's energy and the $Z$ of the daughter nucleus, while the latter, like $\delta_C$, depends in
its evaluation on the details of nuclear structure.  From this equation, it can be seen that
each measured transition establishes an individual value for $\GV$ and, if the CVC assertion is
correct that $\GV$ is not renormalized in the nuclear medium, all such values -- and all the
$\F t$ values themselves -- should be identical within uncertainties, regardless of the specific
nuclei involved.

Our procedure in this paper is to examine all experimental data related to 20 superallowed transitions,
comprising all those that have been well studied, together with others that are now coming under scrutiny
after becoming accessible to precision measurement.  The methods used in data evaluation are presented
in Sec. \ref{s:data}.  The calculations and corrections required to extract $\F t$ values from these data
are described and applied in Sec. \ref{s:Ft}; in the same section, we use the resulting $\F t$ values to
test CVC.  Finally, in Sec. \ref{s:impact} we explore the impact of these results on a number of
weak-interaction issues: CKM unitarity as well as the possible existence of scalar interactions, right-hand
currents and extra $Z$ bosons.  This is much the same pattern as we followed in our last review \cite{HT05}
so we will not describe the formalism again in detail, referring the reader instead to that earlier work.

\section{\label{s:data} Experimental Data}

The $ft$-value that characterizes any $\beta$ transition depends on three measured quantities: the
total transition energy, $Q_{EC}$, the half-life, $t_{1/2}$, of the parent state and the branching ratio,
$R$, for the particular transition of interest.  The $Q_{EC}$-value is required to determine the statistical
rate function, $f$, while the half-life and branching ratio combine to yield the partial half-life, $t$.
In Tables~\ref{QEC}-\ref{reject} we present the measured values of these three quantities and supporting
information for a total of twenty superallowed transitions.

\subsection{\label{eval} Evaluation principles}

In our treatment of the data, we considered all measurements formally published before September 2008 and
those we knew to be in an advanced state of preparation for publication by that date.  We scrutinized all
the original experimental reports in detail.  Where necessary and possible, we used the information provided
there to correct the results for calibration data that have improved since the measurement was made.  If corrections
were evidently required but insufficient information was provided to make them, the results were rejected.  Of
the surviving results, only those with (updated) uncertainties that are within a factor of ten of the most precise
measurement for each quantity were retained for averaging in the tables.  Each datum appearing in the tables is
attributed to its original journal reference {\it via} an alphanumeric code comprising the initial two letters
of the first author's name and the two last digits of the publication date.  These codes are correlated
with the actual reference numbers, \cite{Ad83}-\cite{Zi87}, in Table~\ref{ref}.

The statistical procedures we have followed in analyzing the tabulated data are based on those used by the
Particle Data Group in their periodic reviews of particle properties (e.g. Ref. \cite{PDG08}) and adopted by
us in our previous surveys.  We gave a detailed description of those procedures in our 2004 survey \cite{HT05}
so will not repeat it here.

Our evaluation principles and associated statistical procedures constitute a very conservative approach to
the data.  Unless there is a clearly identifiable reason to reject a result, we include it in our data base
even if it deviates significantly from other measurements of the same quantity, the consequent non-statistical
spread in results being reflected in an increased uncertainty assigned to the average.  Wherever this
occurs, the factor by which the uncertainty has been increased is listed in the ``scale" column of a table.
There are a few examples, though, of a single publication that includes a number of
measurements -- a set of half-lives or $Q_{EC}$ values for example -- most or all of which deviate
substantially from other accepted measurements of the same quantities.  In such cases, we consider that
some systematic problem has been revealed, and exclude all the results from that publication.  If any
measurement with an acceptable uncertainty is nevertheless excluded from our data base, the reason for
its exclusion is listed in Table \ref{reject}.

One particlarly significant, longstanding reference had to be excluded for the first time from this survey.
Our decision to do so deserves a more detailed explanation.  In 1977, Vonach {\it et al.} published in a
single paper \cite{Vo77} the $Q_{EC}$ values for seven superallowed emitters ($^{14}$O, $^{26}$Al$^m$,
$^{34}$Cl, $^{42}$Sc, $^{46}$V, $^{50}$Mn and $^{54}$Co), which they had determined from the $Q$ values for
($^3$He,$t$) reactions on their stable daughters.  They had used a ``precision time-of-flight measuring
system" with the Q3D spectrograph of the Munich Tandem Laboratory to produce uncertainties of 0.4--0.6 keV.
For the time, these were very precise results and consequently they had a major impact on the superallowed
data base for the following three decades.

\begin{figure}[t]
\epsfig{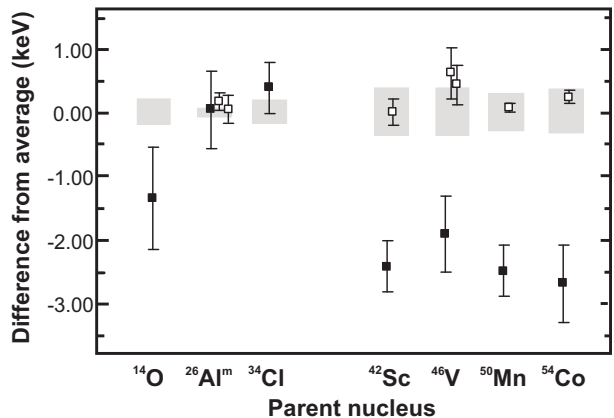}
\caption{Differences between individual measurements and the averages of all measurements for the seven
parent nuclei studied by Vonach {\it et al.}$\,$[Vo77].  The filled squares are the results of the
($^3$He, $t$) measurements of Vonach {\it et al.}; the open squares are recent Penning-trap results
[Sa05, Er06b, Er08, Ge08].  For each parent nucleus, the grey band about the zero line represents the
uncertainty of the average for that case.  Note that all the averages include the results of
Vonach {\it et al.}, the Penning-trap results and any other relevant measurements appearing in
Table \ref{QEC}.}
\label{Vonach}
\end{figure}

The first indication that the Vonach {\it et al.} results might have a problem came with the first
Penning-trap measurement of a superallowed $Q_{EC}$ value \cite{Sa05}.  The new measurement for $^{46}$V
quoted 0.4-keV uncertainty and differed from the old result by 2.4 keV, four of Vonach's claimed standard
deviations.  Within a year, a second Penning-trap measurement \cite{Er06b} had confirmed the new $^{46}$V
result and had also found that the $^{42}$Sc $Q_{EC}$ value differed from the Vonach result by six times the
latter's quoted uncertainty.  Two years later, another Penning-trap measurement \cite{Er08} found the
$^{50}$Mn and $^{54}$Co $Q_{EC}$ values also differed from Vonach's results by a similar amount.  These
most-recent Penning-trap results quoted 0.1-keV uncertainties.  A current overview of the situation for all seven
of the superallowed transitions measured by Vonach appears in Fig.~\ref{Vonach}, where each Vonach result
is compared with the equivalent result(s) from a Penning trap, and both are compared to the average of all
results for the same transition.  With only two transitions, those from $^{26}$Al$^m$ and $^{34}$Cl, showing
agreement and the four cases already mentioned displaying serious disagreement, we believed that the best
approach was to eliminate all the results published in the original Vonach {\it et al.} reference \cite{Vo77}.
This conclusion has been further supported by a very recent ($^3$He,$t$) measurement of the $^{46}$V $Q_{EC}$
value \cite{Fa09} made with much of the same experimental equipment originally used by Vonach
{\it et al.}~thirty years ago.  The new result disagrees with the old measurement and confirms the
new Penning-trap values. 

\begin{figure}[t]
\epsfig{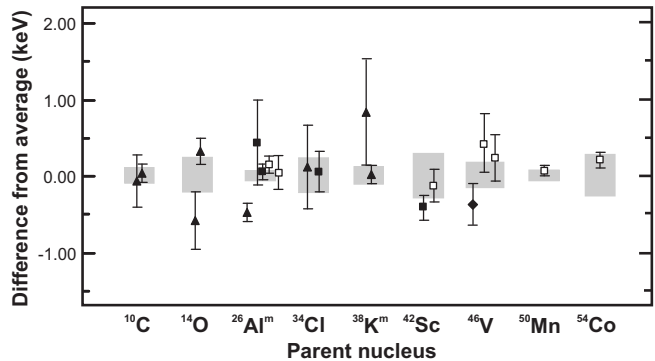}
\caption{For the ``traditional nine" transitions, we plot the differences between individual
measurements and average values; the results of Vonach {\it et al.} [Vo77] have been removed
from the averages.  The open squares are the results of Penning-trap measurements
[Sa05, Er06b, Er08, Ge08]; the filled squares are from combined ($p,\gamma$) and ($n,\gamma$)
measurements [De69 and the references listed in footnotes 6, 7 and 8 of Table \ref{QEC}]; the
filled triangles are from ($p,n$) threshold measurements [Ba84, Ba98, Wh77, To03, Br94,
Ba77c, Ja78, Ha98]; and the filled diamond is the new ($^3$He$,t$) measurement for $^{46}$V
[Fa09].  For each parent nucleus, the grey band about the zero line represents
the uncertainty of the average for that case.}
\label{systQ}
\end{figure}

Before Penning traps could be applied to these measurements, all superallowed $Q_{EC}$ values were
determined {\it via} nuclear reactions.  In addition to those employing ($^3$He,$t$) reactions, two
other types of experiment led to rather precise results: the measurement of ($p,n$) thresholds and the
combined measurements of ($p,\gamma$) and ($n,\gamma$) $Q$-values on the same target, one reaction
leading to the superallowed parent and the other to the daughter.  We are now in a position to compare
the different types of measurement to examine whether there are any systematic differences among them.
A careful study of this issue \cite{Ha06}, restricted to the region around A=26, was undertaken several
years ago and found no evidence of any systematic differences.  We can now confirm this conclusion over
a wider mass range with the help of Fig.~\ref{systQ}.  In that figure we consider nine superallowed
transitions, which we will refer to as the ``traditional nine" cases.  They are the only superallowed
transitions that populate a stable daughter nucleus and, for obvious reasons, were the only ones whose
$Q$ values could be measured to high precision in the pre-trap era.  There are no systematic deviations
apparent in the figure, leading us to conclude that, whatever problems plagued the measurements of Vonach
{\it et al.}$\,$\cite{Vo77}, they were associated with that particular experiment and were not endemic
to a whole class of experiments.  Of course, this conclusion could be strengthened by new Penning-trap
data for $^{10}$C, $^{14}$O, $^{34}$Cl and $^{38}$K$^m$.

\subsection{\label{data} Data Tables}

\begingroup
\squeezetable
\setcounter{LTchunksize}{200}
\setlength{\LTcapwidth}{6.5in}
\begin{center}
\begin{longtable*}{lllllllll}
\caption {Measured results from which the decay transition energies, $Q_{EC}$, have been derived for superallowed
$\beta$-decays.  The lines giving the average superallowed $Q_{EC}$ values themselves are in bold print.
(See Table~\ref{ref} for the correlation between the alphanumeric reference code used in this table and the
actual reference numbers.)
\label{QEC}} \\[-2mm]

\hline\hline
&&&&&&&& \\

\multicolumn{2}{c}{Parent/Daughter}
 & Property\footnotemark[1] 
 & \multicolumn{3}{c}{Measured Energies used to determine $Q_{EC}$ (keV)}
 & \multicolumn{1}{c}{}
 & \multicolumn{2}{c}{Average value} \\[1mm]
\cline{4-6} 
\cline{8-9} 
& & & & & & & &  \\[-2mm]
   \multicolumn{2}{c}{nuclei} & 
 & \multicolumn{1}{c}{1}
 & \multicolumn{1}{c}{2} 
 & \multicolumn{1}{c}{3}
 & \multicolumn{1}{c}{} 
 & \multicolumn{1}{l}{~~~Energy (keV)} 
 & \multicolumn{1}{c}{scale} \\[1mm]

\hline\hline &&&&&&&& \\

\endfirsthead

\multicolumn{9}{l}
{\tablename\ \thetable{} (continued)} \\[2mm]
\hline\hline
\\

\multicolumn{2}{c}{Parent/Daughter}
 & Property\footnotemark[1] 
 & \multicolumn{3}{c}{Measured Energy (keV)}
 & \multicolumn{1}{c}{}
 & \multicolumn{2}{c}{Average value} \\[1mm]
\cline{4-6} 
\cline{8-9} 
& & & & & & & &  \\[-2mm]
   \multicolumn{2}{c}{nuclei} & 
 & \multicolumn{1}{c}{1}
 & \multicolumn{1}{c}{2} 
 & \multicolumn{1}{c}{3}
 & \multicolumn{1}{c}{}  
 & \multicolumn{1}{l}{~~~~Energy (keV)} 
 & \multicolumn{1}{c}{scale} \\[1mm]
\hline
& & & & & & & & \\[-1mm]
\endhead

\hline
\endfoot

\hline\hline
\vspace{-15mm}
\endlastfoot

& & & & & & & \\[-5mm]
$T_z = -1$: & & & & & & & \\
~~ $^{10}$C & $^{10}$B & $Q_{EC}(gs)$ & ~~3647.84 $\pm$ 0.34 ~$\,$[Ba84] & ~~3647.95 $\pm$ 0.12 [Ba98] & & & ~$\:$3647.94 $\pm$ 0.11
  & 1.0   \\
 & & $E_x(d0^+)$ & ~$\;$1740.15 $\pm$ 0.17 ~$\,$[Aj88] & ~~1740.07 $\pm$ 0.02 \footnotemark[2] & & & ~$\:$1740.07 $\pm$ 0.02 & 1.0  \\
 & & $\bm{Q_{EC}(sa)}$ & & & & & {\bf 1907.87 $\pm$ 0.11} & \\[2mm]
~~ $^{14}$O & $^{14}$N & $Q_{EC}(gs)$ & ~$\;$5143.35 $\pm$ 0.60 ~$\,$[Bu61] & ~~5145.09 $\pm$ 0.46 [Ba62] & $\,$5145.57 $\pm$ 0.48 [Ro70]
 & & & \\
 & & & ~$\;$5143.43 $\pm$ 0.37 [Wh77] & ~~5144.34 $\pm$ 0.17 [To03] & & & ~$\:$5144.33 $\pm$ 0.29 & 2.1  \\
 & & $E_x(d0^+)$ & ~$\!$2312.798 $\pm$ 0.011$\:$[Aj91] & & & & 2312.798 $\pm$ 0.011 & \\
 & & $\bm{Q_{EC}(sa)}$ & & & & & {\bf 2831.24 $\pm$ 0.23 \footnotemark[3]} & {\bf 2.3} \\[2mm]
~~ $^{18}$Ne & $^{18}$F & $ME(p)$ & $\:$~~~5316.8 $\pm$ 1.5 ~~$\,$[Ma94] & ~~5317.63 $\pm$ 0.36 [Bl04b] & & & ~$\:$5317.58 $\pm$ 0.35 & 1.0 \\
 & & $ME(d)$ & $\:$~~~873.31 $\pm$ 0.94 ~[Bo64] & ~~~~~875.5 $\pm$ 2.2 ~$\,$[Ho64] & ~~~~876.5 $\pm$ 2.8 ~[Pr67] & & & \\
 & & & ~~~~~877.2 $\pm$ 3.0 ~~$\,$[Se73] & ~~~$\,$873.96 $\pm$ 0.61 [Ro75] & & & ~~~874.02 $\pm$ 0.48 & 1.0 \\
 & & $Q_{EC}(gs)$ & ~~~~~$\;$4438 $\pm$ 9 ~~~~$\:$[Fr63] & & & & ~$\:$4443.54 $\pm$ 0.60 & 1.0 \\
 & & $E_x(d0^+)$ & ~$\;$1041.55 $\pm$ 0.08 ~[Ti95] & & & & ~$\:$1041.55 $\pm$ 0.08 & \\
 & & $\bm{Q_{EC}(sa)}$ & & & & & {\bf 3401.99 $\pm$ 0.60} & \\[2mm]
~~ $^{22}$Mg & $^{22}$Na & $ME(p)$ & ~~~$\;$-401.2 $\pm$ 3.0 ~~$\,$[Ha74c] & ~~~~-400.4 $\pm$ 1.3 ~$\,$\footnotemark[4]
 &  ~~$\;$-400.5 $\pm$ 1.0 ~$\,$[Pa05] & & ~~~~-400.5 $\pm$ 0.8 & 1.0 \\
  & & $ME(d)$ & ~$\:\,$-5184.3 $\pm$ 1.5 ~~$\,$[We68] & ~~$\,$-5182.5 $\pm$ 0.5 ~$\,$[Be68] & ~$\,$-5181.3 $\pm$ 1.7 ~[An70] & & & \\
 & & & ~$\:\,$-5183.2 $\pm$ 1.0 ~~$\,$[Gi72] & ~-5181.56 $\pm$ 0.16 [Mu04] & -5181.08 $\pm$ 0.30
 [Sa04] & & ~-5181.58 $\pm$ 0.19 & 1.7 \\
 & & $Q_{EC}(gs)$ & ~$\;$4781.64 $\pm$ 0.28 ~[Mu04] & ~~4781.40 $\pm$ 0.67 [Sa04] & & & ~$\;$4781.55 $\pm$ 0.25 & 1.0 \\
 & & $E_x(d0^+)$ & ~~~$\,$657.00 $\pm$ 0.14 ~[En98] & & & & ~~~657.00 $\pm$ 0.14 & \\
 & & $\bm{Q_{EC}(sa)}$ & & & & & {\bf 4124.55 $\pm$ 0.28} & \\[2mm] 
~~ $^{26}$Si & $^{26}$Al & $ME(p)$ & ~~~-7145.4 $\pm$ 3.0 ~~~[Ha74c] & ~~$\,$-7139.5 $\pm$ 1.0 ~~$\,$[Pa05] & 
 & & ~~~-7140.1 $\pm$ 1.8  & 1.9 \\
 & & $ME(d0^+)$ & -11981.99 $\pm$ 0.26~$\,$\footnotemark[5] & & & & -11981.99 $\pm$ 0.26 & \\
 & & $\bm{Q_{EC}(sa)}$ & ~~~~~~4850 $\pm$ 13 ~~~[Fr63] & & & & ~~$\,${\bf 4842.0 $\pm$ 1.8} & {\bf 1.0} \\[2mm]
~~ $^{30}$S & $^{30}$P & $ME(p)$ & $\:$~~$\;$-14060 $\pm$ 15 ~~~[Mi67] & ~~~$\,$-14054 $\pm$ 25 ~~$\,$[Mc67] & ~~~-14068 $\pm$ 30 ~~[Ha68] & & & \\ 
 & & & ~-14063.4 $\pm$ 3.0 ~~$\,$[Ha74c] & & & & ~$\:$-14063.1 $\pm$ 2.9& 1.0 \\
 & & $ME(d)$ & ~~~$\,$-20203 $\pm$ 3 ~~~~~[Ha67] &  -20200.58 $\pm$ 0.40 [Re85]& & & -20200.62 $\pm$ 0.40 & 1.0 \\
 & & $Q_{EC}(gs)$ & & & & & ~~~~6137.5 $\pm$ 2.9 & \\
 & & $E_x(d0^+)$ & $\:$~~~677.29 $\pm$ 0.07 $\:$[En98] & & & & ~~~~677.29 $\pm$ 0.07 & \\
 & & $\bm{Q_{EC}(sa)}$ & ~~~~~~5437 $\pm$ 17 ~~~[Fr63] & & & & ~~$\,${\bf 5459.5 $\pm$ 3.9} & {\bf 1.3} \\[2mm]
~~ $^{34}$Ar & $^{34}$Cl & $ME(p)$ & ~-18380.2 $\pm$ 3.0 ~~$\,$[Ha74c] & ~-18378.4 $\pm$ 3.5 ~~$\,$[He01]
 & -18377.10 $\pm$ 0.41$\,$[He02] & & -18377.17 $\pm$ 0.40 & 1.0 \\
 & & $ME(d)$ & $\!$-24440.15 $\pm$ 0.26~$\,$\footnotemark[5] & & & & -24440.15 $\pm$ 0.26 & \\
 & & $\bm{Q_{EC}(sa)}$ & & & & & {\bf ~6062.98 $\pm$ 0.48} & \\[2mm] 
~~ $^{38}$Ca & $^{38}$K & $ME(p)$ & -22058.53 $\pm$ 0.28~$\,$[Ri07] & -22058.01 $\pm$ 0.65 $\,$[Ge07] & & &
 -22058.45 $\pm$ 0.26 & 1.0 \\
 & & $ME(d0^+)$ & -28670.20 $\pm$ 0.32~$\,$\footnotemark[5] & & & & -28670.20 $\pm$ 0.32 & \\
 & & $\bm{Q_{EC}(sa)}$ & & & & & ~{\bf 6611.75 $\pm$ 0.41} & \\[2mm]
~~ $^{42}$Ti & $^{42}$Sc & $ME(p)$ & $\:$~~~-25121 $\pm$ 6 ~~~~$\,$[Mi67] & ~~~-25086 $\pm$ 30 ~~~[Ha68] & ~~~-25124 $\pm$ 13 ~~[Zi72] & & ~$\:$-25120.7 $\pm$ 5.3 & 1.0 \\
 & & $ME(d)$ & -32121.12 $\pm$ 0.29~$\,$\footnotemark[5] & & & & -32121.12 $\pm$ 0.29 & \\
 & & $\bm{Q_{EC}(sa)}$ & & & & & ~~~{\bf 7000.5 $\pm$ 5.4} & \\[2mm]
$T_z = 0$: & & & & & & & & \\
~~ $^{26}$Al$^m$ & $^{26}$Mg & $Q_{EC}(gs)$ & ~$\;$4004.79 $\pm$ 0.55 ~[De69] & ~~4004.41 $\pm$ 0.10 \footnotemark[6]
 & ~~4004.40 $\pm$ 0.22 [Ge08] & & ~~~4004.42 $\pm$ 0.09 & 1.0 \\
 & & $E_x(p0^+)$ & ~$\;$228.305 $\pm$ 0.013$\,$[En98] & & & & ~~~228.305 $\pm$ 0.013 & \\
 & & $\bm{Q_{EC}(sa)}$ & ~~4232.19 $\pm$ 0.12 ~[Br94] & ~~4232.83 $\pm$ 0.13 [Er06b] & & & {\bf ~$\,$4232.66 $\pm$ 0.06 \footnotemark[3]} & {\bf 1.0} \\[2mm]
~~ $^{34}$Cl & $^{34}$S & $\bm{Q_{EC}(sa)}$ & ~~~$\,$5490.3 $\pm$ 1.9 ~~[Ry73a] & ~~~$\,$5491.6 $\pm$ 2.3 ~$\,$[Ha74d] & ~~5491.71 $\pm$ 0.54$\:$[Ba77c]
 & & &  \\
 & & &  ~~5491.65 $\pm$ 0.26 \footnotemark[7] & & & & ~$\:${\bf 5491.64 $\pm$ 0.23} & {\bf 1.0} \\[2mm]
~~ $^{38}$K$^m$ & $^{38}$Ar & $Q_{EC}(gs)$ & ~~5914.82 $\pm$ 0.61 $\,$[Ja78] & & & & ~~~5914.82 $\pm$ 0.61 &  \\
 & & $E_x(p0^+)$ & ~~~~~130.4 $\pm$ 0.3 ~~[En98] & & & & ~~~~~~130.4 $\pm$ 0.3 & \\
 & & $\bm{Q_{EC}(sa)}$ & ~~~$\,$6044.6 $\pm$ 1.5 ~~[Bu79] & ~~6044.38 $\pm$ 0.12 [Ha98] & & & ~$\:${\bf 6044.40 $\pm$ 0.11} & {\bf 1.0} \\[2mm]
~~ $^{42}$Sc & $^{42}$Ca & $\bm{Q_{EC}(sa)}$ & ~~6425.84 $\pm$ 0.17 \footnotemark[8] & ~~6426.13 $\pm$ 0.21 [Er06b] & &
 & ~$\:${\bf 6426.28 $\pm$ 0.30 \footnotemark[3]} &     {\bf 3.0}  \\[2mm]
~~ $^{46}$V  & $^{46}$Ti & $\bm{Q_{EC}(sa)}$ & ~~~$\,$7053.3 $\pm$ 1.8 ~~[Sq76] & ~~7052.90 $\pm$ 0.40 [Sa05]
 & ~~7052.72 $\pm$ 0.31 [Er06b] & & \\
 & & & ~~7052.11 $\pm$ 0.27 $\,$[Fa09] & & & & ~$\:${\bf 7052.49 $\pm$ 0.16} & {\bf 1.3}  \\[2mm]
~~ $^{50}$Mn & $^{50}$Cr & $\bm{Q_{EC}(sa)}$ & ~~7634.48 $\pm$ 0.07 $\,$[Er08] & & & & ~$\:${\bf 7634.45 $\pm$ 0.07 \footnotemark[3]} &  {\bf 1.0} \\[2mm]
~~ $^{54}$Co & $^{54}$Fe & $\bm{Q_{EC}(sa)}$ & ~~8244.54 $\pm$ 0.10 $\,$[Er08] & & & & ~$\:${\bf 8244.37 $\pm$ 0.28 \footnotemark[3]} & {\bf 3.4} \\[2mm]
~~ $^{62}$Ga & $^{62}$Zn & $\bm{Q_{EC}(sa)}$ & ~~9181.07 $\pm$ 0.54 $\:$[Er06a] & & & & ~$\,${\bf 9181.07 $\pm$ 0.54}
 &  \\[2mm]
~~ $^{66}$As & $^{66}$Ge & $ME(p)$ & ~~~$\,$-52018 $\pm$ 30 ~~~[Sc07] & & & & ~~~~~52018 $\pm$ 30 & \\
 & & $ME(d)$ & ~-61607.0 $\pm$ 2.4 $\,$~~[Sc07] & & & & ~~-61607.0 $\pm$ 2.4 & \\
 & & $\bm{Q_{EC}(sa)}$ & ~~~~~~9550 $\pm$ 50 ~~~[Da80] & & & & {\bf ~~~~~9579 $\pm$ 26 } & {\bf 1.0} \\[2mm]
~~ $^{70}$Br & $^{70}$Se & $\bm{Q_{EC}(sa)}$ & ~~~~~~9970 $\pm$ 170$\,$~~[Da80] & & & & {\bf ~~~~~9970 $\pm$ 170}
 &  \\[2mm]
~~ $^{74}$Rb & $^{74}$Kr & $ME(p)$ & ~~~~-51905 $\pm$ 18 ~~$\,$[He02] & ~~-51915.2 $\pm$ 4.0 ~[Ke07] & &
 & ~~-51914.7 $\pm$ 3.9 & 1.0 \\
 & & $ME(d)$ & ~$\,$-62332.0 $\pm$ 2.1 ~~[Ro06] & & & & ~~-62332.0 $\pm$ 2.1 &  \\
 & & $\bm{Q_{EC}(sa)}$ & & & & & ~$\,${\bf 10417.3 $\pm$ 4.4} & \\[2mm]
\vspace{-12mm}
\footnotetext[1]{Abbreviations used in this column are as follows: ``$gs$", transition between ground states;
``$sa$", superallowed transition; ``$p$", parent; ``$d$", daughter; ``$ME$", mass excess; ``$E_x(0^+)$",
excitation energy of the $0^+$ (analog) state.  Thus, for example, ``$Q_{EC}(sa)$" signifies the $Q_{EC}$-value
for the superallowed transition, ``$ME(d)$", the mass excess of the daughter nucleus; and ``$ME(d0^+)$, the mass
excess of the daughter's $0^+$ state.}
\footnotetext[2]{Result based on references [Ba88] and [Ba89].}
\footnotetext[3]{Average result includes the results of $Q_{EC}$ pairs; see Table~\ref{Qdiff}.}
\footnotetext[4]{Result based on references [Bi03], [Se05] and [Je07].}
\footnotetext[5]{Result obtained from data elsewhere in this table.}
\footnotetext[6]{Result based on references [Is80], [Al82], [Hu82], [Be85], [Pr90], [Ki91] and [Wa92].}
\footnotetext[7]{Result based on references [Wa83], [Ra83] and [Li94].}
\footnotetext[8]{Result based on references [Zi87] and [Ki89].}
\end{longtable*}
\end{center}
\endgroup

\pagebreak

\begin{table*}
\caption{$Q_{EC}$-value differences for superallowed $\beta$-decay branches.  These data are also used as input to determine
some of the average $Q_{EC}$-values listed in Table~\ref{QEC}.   (See Table~\ref{ref} for the correlation between the
alphabetical reference code used in this table and the actual reference numbers.)
\label{Qdiff}}
\begin{ruledtabular}
\begin{tabular}{llll}
Parent   
 & \multicolumn{1}{l}{Parent}
 & \multicolumn{2}{c}{$Q_{EC2} - Q_{EC1}$ (keV)} \\[1mm]
\cline{3-4} 
nucleus 1 
 & \multicolumn{1}{l}{nucleus 2}
 & \multicolumn{1}{c}{measurement} 
 & \multicolumn{1}{c}{average\footnotemark[1]} \\
\hline
$^{14}$O & $^{26}$Al$^m$ & 1401.68 $\pm$ 0.13 [Ko87] & 1401.43 $\pm$ 0.24 \\
$^{26}$Al$^m$ & $^{42}$Sc & ~$\,$2193.5 $\pm$ 0.2 ~$\,$[Ko87] & 2193.62 $\pm$ 0.30 \\
$^{42}$Sc & $^{50}$Mn & ~$\,$1207.6 $\pm$ 2.3 ~$\,$[Ha74d] & 1208.17 $\pm$ 0.30 \\
$^{42}$Sc & $^{54}$Co & ~$\,$1817.2 $\pm$ 0.2 ~$\,$[Ko87] & 1818.09 $\pm$ 0.41 \\
$^{50}$Mn & $^{54}$Co & ~$\,$610.09 $\pm$ 0.17 $^{\rm[Ko87]}_{\rm[Ko97b]}$ & ~$\,$609.92 $\pm$ 0.29 \\[-4mm]
\footnotetext[1]{Average values include the results of direct $Q_{EC}$-value measurements: see Table~\ref{QEC}.}
\end{tabular}
\end{ruledtabular}
\end{table*}

\begingroup
\squeezetable
\begin{table*}[b]
\caption{Half-lives, $t_{1/2}$, of superallowed $\beta$-emitters.  (See Table~\ref{ref} for the correlation between the
  alphabetical reference code used in this table and the actual reference numbers.)
\label{t1/2}}
\begin{ruledtabular}
\begin{tabular}{llllllll}
Parent   
 & \multicolumn{4}{c}{Measured half-lives, $t_{1/2}$ (ms)}
 & \multicolumn{1}{c}{}
 & \multicolumn{2}{c}{Average value} \\[1mm]
\cline{2-5} 
\cline{7-8} 
nucleus 
 & \multicolumn{1}{c}{1}
 & \multicolumn{1}{c}{2} 
 & \multicolumn{1}{c}{3} 
 & \multicolumn{1}{c}{4} 
 & \multicolumn{1}{c}{}
 & \multicolumn{1}{c}{$t_{1/2}$ (ms)} 
 & \multicolumn{1}{c}{scale} \\
\hline
 & & & & & & & \\[-4mm]
$T_z = -1$: & & & & & & & \\
~~ $^{10}$C  & ~~$\,$19280 $\pm$ 20 $\,$~~[Az74] & $\:$19295 $\pm$ 15 ~[Ba90] & ~19310 $\pm$ 4 ~~~[Ia08] & & & 19308.0 $\pm$ 3.8  & 1.0   \\
~~ $^{14}$O  & ~~$\,$70480 $\pm$ 150 ~[Al72] & $\:$70588 $\pm$ 28 ~[Cl73] &~70430 $\pm$ 180 [Az74] &~70684 $\pm$ 77 $\,$~[Be78] & & &  \\
 & ~~$\,$70613 $\pm$ 25 $\,$~~[Wi78] & $\:$70560 $\pm$ 49 ~[Ga01] &~70641 $\pm$ 20 $\,$~[Ba04]
 & ~70696 $\pm$ 52 ~$\,$[Bu06] & &~~$\,$70620 $\pm$ 15 & 1.2  \\
~~ $^{18}$Ne & ~~~~1669 $\pm$ 4 ~~~~[Al75] & $\,$~~1687 $\pm$ 9 ~~[Ha75] & $\,$1665.6 $\pm$ 1.9 $\,$[Gr07]  &
             & & $\,$~1667.0 $\pm$ 1.7 & 1.0            \\
~~ $^{22}$Mg & ~~~~3857 $\pm$ 9 ~~~~[Ha75] & 3875.5 $\pm$ 1.2 [Ha03]   & & & & $\,$~3875.2 $\pm$ 2.4 & 2.0 \\
~~ $^{26}$Si & ~~~~2210 $\pm$ 21 $\,$~~[Ha75] & ~~2240 $\pm$ 10 ~[Wi80] & 2228.3 $\pm$ 2.7 ~[Ma08] & & & ~$\,$2228.8 $\pm$ 2.9 & 1.1 \\
~~ $^{30}$S  & ~~~~1180 $\pm$ 40 $\,$~~[Ba67] & ~~1220 $\pm$ 30 ~[Mo71] &1178.3 $\pm$ 4.8 ~[Wi80] &   & & $\,$~1179.4 $\pm$ 4.7 & 1.0 \\
~~ $^{34}$Ar & ~~~844.5 $\pm$ 3.4 ~~[Ha74a] & ~843.8 $\pm$ 0.4 $\,$[Ia06]  & & & & ~~~843.8 $\pm$ 0.4 & 1.0 \\
~~ $^{38}$Ca & ~~~~~$\,$470 $\pm$ 20 $\:$~~[Ka68] & ~~~\,439 $\pm$ 12 $\,$~[Ga69] & ~~~~450 $\pm$ 70 ~~[Zi72] &~~~~430 $\pm$ 12 $\,$~[Wi80]
             & & ~~~440.0 $\pm$ 7.8  & 1.2          \\
~~ $^{42}$Ti & ~~~~~$\,$200 $\pm$ 20 $\:$~~[Ni69] & ~~~\,202 $\pm$ 5 ~~~[Ga69] & ~~~~173 $\pm$ 14 ~~[Al69] & & & ~~~198.8 $\pm$ 6.3 & 1.4 \\[1mm]
$T_z = 0$: & & & & & & &  \\
~~ $^{26}$Al$^m$ &  ~~~~6346 $\pm$ 5 $\,$~~~~[Fr69a] & $\,$~~6346 $\pm$ 5 $\,$~~~[Az75] & 6339.5 $\pm$ 4.5 $\,$~[Al77] & 6346.2 $\pm$ 2.6 ~[Ko83] & &    $\,$~6345.0 $\pm$ 1.9    & 1.0      \\
  & ~~~~6345 $\pm$ 14 ~~~[Sc05] & & & & & & \\
~~ $^{34}$Cl & ~~~~1526 $\pm$ 2 $\,$~~~~[Ry73a] & 1525.2 $\pm$ 1.1 $\,$~[Wi76] & 1527.7 $\pm$ 2.2 $\,$~[Ko83]
 & 1526.8 $\pm$ 0.5 ~[Ia06] & & 1526.55 $\pm$ 0.44   &  1.0   \\
~~ $^{38}$K$^m$ & ~~~925.6 $\pm$ 0.7 $\,$~~[Sq75] & $\,$~922.3 $\pm$ 1.1 $\,$~[Wi76] & 921.71 $\pm$ 0.65 [Wi78]
 & 924.15 $\pm$ 0.31$\,$[Ko83] & &  &  \\
 & ~~~924.4 $\pm$ 0.6 $\,$~~[Ba00] & 924.46 $\pm$ 0.14 [Ba08] &  & & & $\,$~924.33 $\pm$ 0.27 & 2.3  \\
~~ $^{42}$Sc &  ~$\,$680.98 $\pm$ 0.62 ~[Wi76] & 680.67 $\pm$ 0.28 [Ko97a]  &    &  &  & $\,$~680.72 $\pm$ 0.26  &  1.0      \\
~~ $^{46}$V &  ~$\,$422.47 $\pm$ 0.39 ~[Al77] & 422.28 $\pm$ 0.23 [Ba77a] & 422.57 $\pm$ 0.13 [Ko97a]  &   &  & $\,$~422.50 $\pm$ 0.11 & 1.0    \\
~~ $^{50}$Mn &  ~~~284.0 $\pm$ 0.4 $\,$~~[Ha74b] & $\,$~282.8 $\pm$ 0.3 $\,$~[Fr75] & 282.72 $\pm$ 0.26 [Wi76]
 & 283.29 $\pm$ 0.08 [Ko97a] & & &   \\
 & ~$\,$283.10 $\pm$ 0.14 ~[Ba06] & & & & & $\,$~283.21 $\pm$ 0.11 & 1.7  \\
~~ $^{54}$Co & $\,$~~$\,$193.4 $\pm$ 0.4 $\,$~~[Ha74b] & $\,$~193.0 $\pm$ 0.3 $\,$~[Ho74] & 193.28 $\pm$ 0.18 [Al77]
 & 193.28 $\pm$ 0.07 [Ko97a] & & 193.271 $\pm$ 0.063 & 1.0        \\
~~ $^{62}$Ga & ~$\,$115.84 $\pm$ 0.25 ~[Hy03] & 116.19 $\pm$ 0.04 [Bl04a] & 116.09 $\pm$ 0.17 [Ca05]
 & 116.01 $\pm$ 0.19 [Hy05] & & &         \\
             & 116.100 $\pm$ 0.025  [Gr08] & & & & & 116.121 $\pm$ 0.040 & 1.9    \\
~~ $^{66}$As &  ~~~95.78 $\pm$ 0.39 ~[Al78] & $\,$~95.77 $\pm$ 0.28 [Bu88] & ~~~~~97 $\pm$ 2 ~~~~[Ji02]  &  &
 & ~~~95.79 $\pm$ 0.23 & 1.0         \\
~~ $^{70}$Br &  ~~~~$\,$80.2 $\pm$ 0.8 $\,$~~[Al78] & $\,$~78.54 $\pm$ 0.59 [Bu88] &    &  &  & ~~~79.12 $\pm$ 0.79 & 1.7       \\
~~ $^{74}$Rb &  ~~~64.90 $\pm$ 0.09 ~[Oi01] & 64.761 $\pm$ 0.031$\,$[Ba01] &    &   &  & $\,$~64.776 $\pm$ 0.043 &  1.5   \\

\end{tabular}
\end{ruledtabular}
\end{table*}
\endgroup

\begingroup
\squeezetable
\begin{table*}
\caption{Measured results from which the branching ratios, R, have been derived for superallowed $\beta$-transitions.
The lines giving the average superallowed branching ratios themselves are in bold print. ( See Table~\ref{ref} for
the correlation between the alphabetical reference code used in this table and the actual reference numbers.)
\label{R}}
\begin{ruledtabular}
\begin{tabular}{llllllll}
\multicolumn{2}{c}{Parent/Daughter}
 & Daughter state
 & \multicolumn{2}{c}{Measured Branching Ratio, R (\%)}
 & \multicolumn{1}{c}{}
 & \multicolumn{2}{c}{Average value} \\[1mm]
\cline{4-6} 
\cline{7-8} 
   \multicolumn{2}{c}{nuclei}
 & \multicolumn{1}{c}{$E_x$ (MeV)}
 & \multicolumn{1}{c}{1}
 & \multicolumn{1}{c}{2} 
 & \multicolumn{1}{c}{} 
 & \multicolumn{1}{c}{R (\%)} 
 & \multicolumn{1}{c}{scale} \\

\hline
& & & & & & & \\[-4mm]
 $T_z = -1$: & & & & & & & \\
~~ $^{10}$C & $^{10}$B & 2.16 & 0$^{+0.0008}_{-0}$ [Go72] & & & 0$^{+0.0008}_{-0}$ &   \\
 & & {\bf 1.74} & 1.468 $\pm$ 0.014 [Ro72]& 1.473 $\pm$ 0.007 [Na91] & & & \\
 & & & 1.465 $\pm$ 0.009 [Kr91] & 1.4625 $\pm$ 0.0025 [Sa95] & & & \\
 & & & 1.4665 $\pm$ 0.0038 [Fu99] & & & {\bf 1.4646 $\pm$ 0.0019} & {\bf 1.0} \\
~~ $^{14}$O & $^{14}$N & gs & 0.68 $\pm$ 0.10 [Sh55,To05] & 0.74 $\pm$ 0.05 [Fr63,To05]] & & & \\
 & & & 0.54 $\pm$ 0.02 [Si66,To05] & & & 0.571 $\pm$ 0.068 & 3.7 \\
 & & 3.95 & 0.062 $\pm$ 0.007 [Ka69] & 0.058 $\pm$ 0.004 [Wi80] & & & \\
 & & & 0.053 $\pm$ 0.002 [He81] & & & 0.0545 $\pm$ 0.0019 & 1.1 \\
 & & {\bf 2.31} & & & & {\bf 99.374 $\pm$ 0.068} & \\
~~ $^{18}$Ne & $^{18}$F & {\bf 1.04} & 9 $\pm$ 3 [Fr63] & 7.70 $\pm$ 0.21\footnotemark[1] [Ha75] & & {\bf 7.70 $\pm$ 0.21} & {\bf 1.0} \\
~~ $^{22}$Mg & $^{22}$Na & {\bf 0.66} & 54.0 $\pm$ 1.1 [Ha75] & 53.15 $\pm$ 0.12 [Ha03] & & {\bf 53.16 $\pm$ 0.12} & {\bf 1.0} \\
~~ $^{26}$Si & $^{26}$Al & 1.06 & 21.8 $\pm$ 0.8 [Ha75] & 21.21 $\pm$ 0.64 [Ma08] & & 21.4 $\pm$ 0.5 & 1.0 \\
 & & {\bf 0.23} & & & & {\bf 75.49 $\pm$ 0.57\footnotemark[1]} & \\
~~ $^{30}$S & $^{30}$P & gs & 20 $\pm$ 1 [Fr63] & & & 20 $\pm$ 1 &  \\
 & & {\bf 0.68} & & & & {\bf 77.4 $\pm$ 1.0\footnotemark[1]} & \\
~~ $^{34}$Ar & $^{34}$Cl & 0.67 & 2.49 $\pm$ 0.10 [Ha74a] & & & 2.49 $\pm$ 0.10 & \\
 & & {\bf gs} & & & & {\bf 94.45 $\pm$ 0.25\footnotemark[1]} & \\
~~ $^{42}$Ti & $^{42}$Sc & 0.61 & 56 $\pm$ 14 [Al69] & & & 56 $\pm$ 14 & \\
 & & {\bf gs} & & & & {\bf 43 $\pm$ 14\footnotemark[1]} & \\[1mm]
 $T_z = 0$: & & & & & & &  \\
~~ $^{26}$Al$^m$ & $^{26}$Mg & {\bf gs} & $>$99.997 [Ki91] & & & {\bf 100.000}$\bm{^{+0}_{-0.003}}$ & \\
~~ $^{34}$Cl & $^{34}$S & {\bf gs} & $>$99.988 [Dr75] & & & {\bf 100.000}$\bm{^{+0}_{-0.012}}$ & \\
~~ $^{38}$K$^m$ & $^{38}$Ar & 3.38 & $<$0.0019 [Ha94] & $<$0.0008 [Le08] & & 0$^{+0.0008}_{-0}$ & \\
 & & gs($^{38}$K) & 0.0330 $\pm$ 0.0043 [Le08] & & & 0.0330 $\pm$ 0.0043  & \\
 & & {\bf gs} & & & & {\bf 99.9670} $\bm{^{+0.0043}_{-0.0044}}$ & \\
~~ $^{42}$Sc & $^{42}$Ca & 1.84 & 0.0063 $\pm$ 0.0026 [In77] & 0.0022 $\pm$ 0.0017 [De78] & & & \\
 & & & 0.0103 $\pm$ 0.0031 [Sa80] & 0.0070 $\pm$ 0.0012 [Da85] & & 0.0059 $\pm$ 0.0014 & 1.6 \\
  & & {\bf gs} & & & & {\bf 99.9941 $\pm$ 0.0014} & \\
~~$^{46}$V & $^{46}$Ti & 2.61 & 0.0039 $\pm$ 0.0004 [Ha94] & & & 0.0039 $\pm$ 0.0004 & \\
 & & 4.32 & 0.0113 $\pm$ 0.0012 [Ha94] & & & 0.0113 $\pm$ 0.0012 & \\
 & & $\Sigma$GT\footnotemark[2] & $<$0.004 & & & 0$^{+0.004}_{-0}$ & \\
 & & {\bf gs} & & & & {\bf 99.9848}$\bm{^{+0.0013}_{-0.0042}}$ & \\
~~$^{50}$Mn & $^{50}$Cr & 3.63 & 0.057 $\pm$ 0.003 [Ha94] & & & 0.057 $\pm$ 0.003 & \\
 & & 3.85 & $<$0.0003 [Ha94] & & & $0^{+0.0003}_{-0}$ & \\
 & & 5.00 & 0.0007 $\pm$ 0.0001 [Ha94] & & & 0.0007 $\pm$ 0.0001 & \\
 & & {\bf gs} & & & & {\bf 99.9423 $\pm$ 0.0030} & \\
~~$^{54}$Co & $^{54}$Fe & 2.56 & 0.0045 $\pm$ 0.0006 [Ha94] & & & 0.0045 $\pm$ 0.0006 & \\
 & & $\Sigma$GT\footnotemark[2] & $<$0.03 & & & 0$^{+0.03}_{-0}$ & \\
 & & {\bf gs} & & & & {\bf 99.9955}$\bm{^{+0.0006}_{-0.0300}}$ & \\
~~$^{62}$Ga & $^{62}$Zn & $\Sigma$GT\footnotemark[2] & 0.142 $\pm$ 0.008 [Fi08] & 0.107 $\pm$ 0.024 [Be08] &
 & 0.139 $\pm$ 0.011 & 1.4 \\
 & & {\bf gs} & & & & {\bf 99.862 $\pm$ 0.011} & \\
~~$^{74}$Rb & $^{74}$Kr & $\Sigma$GT\footnotemark[2] & 0.50 $\pm$ 0.10 [Pi03] & & & 0.50 $\pm$ 0.10 & \\
 & & {\bf gs} & & & & {\bf 99.50 $\pm$ 0.10} & \\[-4mm]

\footnotetext[1]{Result also incorporates data from Table~\ref{BDG}}
\footnotetext[2]{designates total Gamow-Teller transitions to levels not explicitly listed; values were derived
with the help of calculations in [Ha02].}
\end{tabular}
\end{ruledtabular}
\end{table*}
\endcenter
\endgroup

\begingroup
\squeezetable
\begin{table*}
\caption{Relative intensities of $\beta$-delayed $\gamma$-rays in the superallowed $\beta$-decay daughters.  These data
are used to determine some of the branching ratios presented in Table~\ref{R}.  (See Table~\ref{ref}
for the correlation between the alphabetical reference code used in this table and the actual reference numbers.)
\label{BDG}}
\begin{ruledtabular}
\begin{tabular}{llllllll}
\multicolumn{2}{c}{Parent/Daughter}
 & \multicolumn{1}{c}{daughter}
 & \multicolumn{2}{c}{Measured $\gamma$-ray Ratio}
 & \multicolumn{1}{c}{}
 & \multicolumn{2}{c}{Average value} \\[1mm]
\cline{4-6} 
\cline{7-8} 
   \multicolumn{2}{c}{nuclei}
 & \multicolumn{1}{c}{ratios\footnotemark[1]}
 & \multicolumn{1}{c}{1}
 & \multicolumn{1}{c}{2} 
 & \multicolumn{1}{c}{} 
 & \multicolumn{1}{c}{Ratio} 
 & \multicolumn{1}{c}{scale} \\

\hline
& & & & & & & \\[-4mm]
$^{18}$Ne & $^{18}$F & $\gamma_{660}/\gamma_{1042}$ & ~~~0.021 $\pm$ 0.003 ~~~[Ha75] & 0.0169 $\pm$ 0.0004 [He82] & & & \\
 & & & ~$\,$0.0172 $\pm$ 0.0005 ~$\,$[Ad83]& & & 0.0171 $\pm$ 0.0003 & 1.0 \\
$^{26}$Si & $^{26}$Al & $\gamma_{1622}/\gamma_{829}$ & ~~~0.149 $\pm$ 0.016 ~~~[Mo71] & ~$\,$0.134 $\pm$ 0.005 ~$\,$[Ha75]& & & \\
 & & & ~$\,$0.1245 $\pm$ 0.0023 ~$\,$[Wi80]& 0.1301 $\pm$ 0.0062 [Ma08] & & 0.1269 $\pm$ 0.0026 & 1.3 \\
 & & $\gamma_{1655}/\gamma_{829}$ & 0.00145 $\pm$ 0.00032 [Wi80] & & & 0.0015 $\pm$ 0.0003 & \\
 & & $\gamma_{1843}/\gamma_{829}$ & ~~~0.013 $\pm$ 0.003 ~~~[Mo71] & ~$\,$0.016 $\pm$ 0.003 ~$\,$[Ha75] & & & \\
 & & & 0.01179 $\pm$ 0.00027 [Wi80] & & & 0.0118 $\pm$ 0.0003 & 1.0 \\
 & & $\gamma_{2512}/\gamma_{829}$ &0.00282 $\pm$ 0.00010 [Wi80] & & & 0.0028 $\pm$ 0.0001 & \\
 & & $\gamma_{\rm total}/\gamma_{829}$ & & & & 0.1430 $\pm$ 0.0026 & \\
$^{30}$S & $^{30}$P & $\gamma_{709}/\gamma_{677}$ & ~~~0.006 $\pm$ 0.003 ~~~[Mo71] & 0.0037 $\pm$ 0.0009 [Wi80] & & 0.0039 $\pm$ 0.0009 & 1.0 \\
 & & $\gamma_{2341}/\gamma_{677}$ & ~~~0.033 $\pm$ 0.002 ~~~[Mo71] & 0.0290 $\pm$ 0.0006 [Wi80] & & 0.0293 $\pm$ 0.0011 & 1.9 \\
 & & $\gamma_{3019}/\gamma_{677}$ & 0.00013 $\pm$ 0.00006 [Wi80] & & & 0.0001 $\pm$ 0.0001 & \\
 & & $\gamma_{\rm total}/\gamma_{677}$ & & & & 0.0334 $\pm$ 0.0014 & \\
$^{34}$Ar & $^{34}$S & $\gamma_{461}/\gamma_{666}$ & ~~~~$\,$0.28 $\pm$ 0.16 ~~~~$\:$[Mo71] & ~$\,$0.365 $\pm$ 0.036 ~$\,$[Ha74a] & & ~$\,$0.361 $\pm$ 0.035 & 1.0 \\
 & & $\gamma_{2580}/\gamma_{666}$ & ~~~~$\,$0.38 $\pm$ 0.09 ~~~~$\:$[Mo71] & ~$\,$0.345 $\pm$ 0.01 ~~~[Ha74a] & & ~$\,$0.345 $\pm$ 0.010 & 1.0 \\
 & & $\gamma_{3129}/\gamma_{666}$ & ~~~~$\,$0.67 $\pm$ 0.08 ~~~~$\:$[Mo71] & ~$\,$0.521 $\pm$ 0.012 ~$\,$[Ha74a] & & ~$\,$0.524 $\pm$ 0.022 & 1.8 \\
 & & $\gamma_{\rm total}/\gamma_{666}$ & & & & ~$\,$1.231 $\pm$ 0.043 & \\
$^{42}$Ti & $^{42}$Sc & $\gamma_{2223}/\gamma_{611}$ & ~~~0.012 $\pm$ 0.004 ~~~[Ga69] & & & ~$\,$0.012 $\pm$ 0.004 & \\
 & & $\gamma_{\rm total}/\gamma_{611}$ & ~~~0.023 $\pm$ 0.012 [Ga69,En90] & & & ~$\,$0.023 $\pm$ 0.012 & \\

\footnotetext[1]{$\gamma$-ray intensities are denoted by $\gamma_{E}$, where $E$ is the $\gamma$-ray energy in keV.}
\end{tabular}
\end{ruledtabular}
\end{table*}
\endgroup

\begingroup
\squeezetable
\begin{table*}
\caption{References for which the original decay-energy results have been updated to incorporate the most recent calibration standards.  (See Table~\ref{ref} for the correlation between the alphabetical reference code used in this table and the actual reference numbers.)
\label{update}}
\vskip 1mm
\begin{ruledtabular}
\begin{tabular}{lll}
  \multicolumn{1}{l}{References (parent nucleus)\footnotemark[1]}
& \multicolumn{1}{l}{}
& \multicolumn{1}{l}{Update procedure} \\
\hline
\\[-3mm]
\textbullet~Bo64$\,$($^{18}$Ne), Ba84$\,$($^{10}$C), Br94$\,$($^{26}$Al$^m$) &~~~~ & \textbullet~We have converted all original (p,n) threshold measurements to $Q$-values \\ 
 Ba98$\,$($^{10}$C), Ha98$\,$($^{38}$K$^m$), To03$\,$($^{14}$O) & &  using the most recent mass excesses [Au03]. \\[1mm]
\textbullet~Ry73a$\,$($^{34}$Cl), Sq76$\,$($^{46}$V), Ba77c$\,$($^{34}$Cl) & & \textbullet~These (p,n) threshold measurements have been adjusted to reflect recent \\
Wh77$\,$($^{14}$O) & & calibration $\alpha$-energies [Ry91] before being converted to $Q$-values. \\[1mm]
\textbullet~Pr67$\,$($^{18}$Ne) & & \textbullet~Before conversion to a $Q$-value, this (p,n) threshold was adjusted to reflect a  \\
 & & new value for the $^7$Li(p,n) threshold [Wh85], which was used as calibration. \\[1mm]
\textbullet~Ja78$\,$($^{38}$K$^m$) & & \textbullet~This (p,n) threshold was measured relative to those for $^{10}$C and $^{14}$O; we have \\
 & & adjusted it based on average $Q$-values obtained for those decays in this work. \\[1mm]
\textbullet~Bu79$\,$($^{38}$K$^m$) & & \textbullet~Before conversion to a $Q$-value, this (p,n) threshold was adjusted to reflect the  \\
 & & modern value for the $^{35}$Cl(p,n) threshold [Au03], which was used as calibration. \\[1mm]
\textbullet~Bu61$\,$($^{14}$O), Ba62$\,$($^{14}$O) & & \textbullet~These $^{12}$C($^3$He,n) threshold measurements have been adjusted for updated \\
 & & calibration reactions based on current mass excesses [Au03]. \\[1mm]
\textbullet~Ha74d$\,$($^{34}$Cl) & & \textbullet~These ($^3$He,t) reaction $Q$-values were calibrated by the $^{27}$Al($^3$He,t) reaction \\
 & & to excited states in $^{27}$Si; they have been revised according to modern mass \\
 & & excesses [Au03] and excited-state energies [En98]. \\[1mm]
\textbullet~Ba88 and Ba89$\,$($^{10}$C) & & \textbullet~These measurements of excitation energies in $^{10}$B have
been updated to \\
 & & modern $\gamma$-ray standards [He00]. \\[1mm] 
\textbullet~Ki89$\,$($^{42}$Sc) & & \textbullet~This $^{41}$Ca(p,$\gamma$) reaction $Q$-value was measured relative to that for $^{40}$Ca(p,$\gamma$); \\
 & & we have slightly revised the result based on modern mass excesses [Au03]. \\[1mm]
\textbullet~Ha74c$\,$($^{22}$Mg, $^{26}$Si, $^{30}$S, $^{34}$Ar) & & \textbullet~These (p,t) reaction $Q$-values have been adjusted to reflect the current $Q$- \\
 & & value for the $^{16}$O(p,t) reaction [Au03], against which they were calibrated. \\[-4mm]    

\footnotetext[1]{These references all appear in Table~\ref{QEC} under the appropriate parent nucleus.}
\end{tabular}
\end{ruledtabular}
\end{table*}
\endgroup

The $Q_{EC}$-value data appear in Tables~\ref{QEC} and \ref{Qdiff}.  For the ``traditional nine"
superallowed decays -- those of $^{10}$C, $^{14}$O, $^{26}$Al$^m$, $^{34}$Cl, $^{38}$K$^m$, $^{42}$Sc,
$^{46}$V, $^{50}$Mn and $^{54}$Co -- with stable daughter nuclei, their $Q_{EC}$ values were all
determined in the past by direct reaction measurements of that property.  More recently, a growing number of
Penning-trap measurements, also extending to nuclei outside of the traditional nine, determine the parent and
daughter masses in a single experiment, thus effectively measuring the $Q_{EC}$ value directly.
Measurements of both types are identified in column 3 of Table~\ref{QEC} by ``$Q_{EC}(sa)$" and
each individual result is itemized with its appropriate reference in the next three columns.  The
weighted average of all measurements for a particular decay appears in column 7, with the corresponding
scale factor (see Sec.~\ref{eval}) in column 8.  A few of these cases, like $^{34}$Cl and all the cases
from $^{42}$Sc to $^{62}$Ga, have no further complications.  There are other cases, however, in which
$Q_{EC}$-value differences have been measured in addition to the individual $Q_{EC}$-values.  These
measurements are presented in Table~\ref{Qdiff}.  They have been dealt with in combination with the
direct $Q_{EC}$-value measurements, as described in Ref.~\cite{HT05}, with the final average $Q_{EC}$
value appearing in column 7 of Table~\ref{QEC} and the average difference in column 4 of Table~\ref{Qdiff}.
Both are flagged with footnotes to indicate the interconnection.

There are two cases, $^{26}$Al$^m$ and $^{38}$K$^m$, in which the superallowed decay originates from
an isomeric state.  For both, there are $Q_{EC}$-value measurements that correspond to the ground state
as well as to the isomer. Obviously, these two sets of measurements are simply related to one another
by the excitation energy of the isomeric state in the parent.  In Table~\ref{QEC} the set of
measurements for the ground-state $Q_{EC}$-value and for the excitation energy of the isomeric state
appear in separate rows, each with its identifying property given in column 3 and its weighted average
appearing in column 7.  In the row below, the average value given in column 7 for the superallowed
transition is the weighted average not only of the direct superallowed $Q_{EC}$-value measurements
in that row, but also of the result derived from the two preceeding rows.  Note that in all cases
the $Q_{EC}$-value for the superallowed transition appears in bold-face type.

\begingroup
\squeezetable
\begin{table*}
\caption{References from which some or all results have been rejected even though their quoted uncertainties
qualified them for inclusion.  (See Table~\ref{ref} for the correlation between the alphabetical reference
code used in this table and the actual reference numbers.)
\label{reject}}
\vskip 1mm
\begin{ruledtabular}
\begin{tabular}{llll}
  \multicolumn{2}{l}{References (parent nucleus)}
& \multicolumn{1}{l}{}
& \multicolumn{1}{l}{Reason for rejection} \\
\hline
1. & Decay-energies: &~~~~ & \\[1mm]
 & \textbullet~Pa72$\,$($^{30}$S) & & \textbullet~No calibration is given for the measured (p,t) reaction $Q$-values; update \\
 & & & is clearly required but none is possible. \\[1mm]
 & \textbullet~No74$\,$($^{22}$Mg) & & \textbullet~Calibration reaction $Q$-values have changed but calibration process is too \\
 & & & complex to update. \\[1mm]
 & \textbullet~Ro74$\,$($^{10}$C) & & \textbullet~P.H. Barker (co-author) later considered that inadequate attention had \\
 & & & been paid to target surface purity [Ba84]. \\[1mm]
 & \textbullet~Ba77b$\,$($^{10}$C) & & \textbullet~P.H. Barker (co-author) later stated [Ba84] that the (p,t) reaction $Q$-value \\
 & & & could not be updated to incorporate modern calibration standards. \\[1mm]
 & \textbullet~Vo77$\,$($^{14}$O, $^{26}$Al$^m$, $^{34}$Cl, $^{42}$Sc, $^{46}$V, $^{50}$Mn, $^{54}$Co)
 & & \textbullet~Most of the results in this reference disagree significantly with more recent \\
 & & & and accurate measurements.  Our justification for rejection is presented in \\
 & & & more detail in the text. \\[1mm]
 & \textbullet~Wh81 and Ba98$\,$($^{14}$O) & & \textbullet~The result in [Wh81] was updated in [Ba98] but then eventually withdrawn \\
 & & & by P.H. Barker (co-author) in [To03]. \\[2mm]
2. & Half-lives: & & \\[1mm]
 & \textbullet~Ja60$\,$($^{26}$Al$^m$), He61$\,$($^{14}$O), Ba62$\,$($^{14}$O),  & & \textbullet~Quoted uncertainties are too small, and results likely biased, in light of  \\
 & Fr63$\,$($^{14}$O), Fr65b$\,$($^{42}$Sc, $^{46}$V, $^{50}$Mn) & &  statistical difficulties more recently understood (see [Fr69a]).  In particular, \\
 & Si72$\,$($^{14}$O) & & ``maximum-likelihood" analysis was not used. \\[1mm]
 & \textbullet~Ha72a$\,$($^{26}$Al$^m$, $^{34}$Cl, $^{38}$K$^m$, $^{42}$Sc) & & \textbullet~All four quoted half-lives are systematically higher than more recent and \\
 & & & accurate measurements. \\[1mm]
 & \textbullet~Ro74$\,$($^{10}$C) & & \textbullet~P.H. Barker (co-author) later considered that pile-up had been \\
 & & & inadequately accounted for [Ba90]. \\[1mm]
 & \textbullet~Ch84$\,$($^{38}$K$^m$) & & \textbullet~``Maximum-likelihood" analysis was not used.  \\[2mm]
3. & Branching-ratios: & & \\[1mm]
 & \textbullet~Fr63$\,$($^{26}$Si)& & \textbullet~Numerous impurities present; result is obviously wrong. \\

\end{tabular}
\end{ruledtabular}
\end{table*}
\endgroup

\begingroup
\squeezetable
\begin{table*}
\caption{Reference key, relating alphabetical reference codes used in Tables~\ref{QEC}-\ref{reject} to the actual reference numbers.
\label{ref}}
\vskip 1mm
\begin{ruledtabular}
\begin{tabular}{llllllllllll}
  Table & Reference & Table & Reference & Table & Reference & Table & Reference & Table & Reference & Table & Reference \\ 
  code & number & code & number & code & number & code & number & code & number & code & number \\
\hline
  Ad83  & \cite{Ad83}  &
  Aj88  & \cite{Aj88}  &
  Aj91  & \cite{Aj91}  &
  Al69  & \cite{Al69}  &
  Al72  & \cite{Al72}  &
  Al75  & \cite{Al75}  \\
  Al77  & \cite{Al77}  &
  Al78  & \cite{Al78}  &
  Al82  & \cite{Al82}  &
  An70  & \cite{An70}  &
  Au03  & \cite{Au03}  &
  Az74  & \cite{Az74}  \\
  Az75  & \cite{Az75}  &
  Ba62  & \cite{Ba62}  &
  Ba67  & \cite{Ba67}  &
  Ba77a & \cite{Ba77a} &
  Ba77b & \cite{Ba77b} &
  Ba77c & \cite{Ba77c} \\
  Ba84  & \cite{Ba84}  &
  Ba88  & \cite{Ba88}  &
  Ba89  & \cite{Ba89}  &
  Ba90  & \cite{Ba90}  &
  Ba98  & \cite{Ba98}  &
  Ba00  & \cite{Ba00}  \\
  Ba01  & \cite{Ba01}  &
  Ba04  & \cite{Ba04}  &
  Ba06  & \cite{Ba06}  &
  Ba08  & \cite{Ba08}  &
  Be68  & \cite{Be68}  &
  Be78  & \cite{Be78}  \\
  Be85  & \cite{Be85}  &
  Be08  & \cite{Be08}  &
  Bi03  & \cite{Bi03}  &
  Bl04a & \cite{Bl04a} &
  Bl04b & \cite{Bl04b} &
  Bo64  & \cite{Bo64}  \\
  Br94  & \cite{Br94}  &
  Bu61  & \cite{Bu61}  &
  Bu79  & \cite{Bu79}  &
  Bu88  & \cite{Bu88}  &
  Bu06  & \cite{Bu06}  &
  Ca05  & \cite{Ca05}  \\
  Ch84  & \cite{Ch84}  &
  Cl73  & \cite{Cl73}  &
  Da80  & \cite{Da80}  &
  Da85  & \cite{Da85}  &
  De69  & \cite{De69}  &
  De78  & \cite{De78}  \\
  Dr75  & \cite{Dr75}  &
  En90  & \cite{En90}  &
  En98  & \cite{En98}  &
  Er06a & \cite{Er06a} &
  Er06b & \cite{Er06b} &
  Er08  & \cite{Er08}  \\
  Fa09  & \cite{Fa09}  &
  Fi08  & \cite{Fi08}  &
  Fr63  & \cite{Fr63}  &
  Fr65b & \cite{Fr65b} &
  Fr69a & \cite{Fr69a} &
  Fr75  & \cite{Fr75}  \\
  Fu99  & \cite{Fu99}  &
  Ga69  & \cite{Ga69}  &
  Ga01  & \cite{Ga01}  &
  Ge07  & \cite{Ge07}  &
  Ge08  & \cite{Ge08}  &
  Gi72  & \cite{Gi72}  \\
  Go72  & \cite{Go72}  &
  Gr07  & \cite{Gr07}  &
  Gr08  & \cite{Gr08}  &
  Ha67  & \cite{Ha67}  &
  Ha68  & \cite{Ha68}  &
  Ha72a & \cite{Ha72a} \\
  Ha74a & \cite{Ha74a} &
  Ha74b & \cite{Ha74b} &
  Ha74c & \cite{Ha74c} &
  Ha74d & \cite{Ha74d} &
  Ha75  & \cite{Ha75}  &
  Ha94  & \cite{Ha94}  \\
  Ha98  & \cite{Ha98}  &
  Ha02  & \cite{Ha02}  &
  Ha03  & \cite{Ha03}  &
  He61  & \cite{He61}  &
  He81  & \cite{He81}  & 
  He82  & \cite{He82}  \\
  He00  & \cite{He00}  &
  He01  & \cite{He01}  &
  He02  & \cite{He02}  &
  Ho64  & \cite{Ho64}  &
  Ho74  & \cite{Ho74}  &
  Hu82  & \cite{Hu82}  \\
  Hy03  & \cite{Hy03}  &
  Hy05  & \cite{Hy05}  &
  Ia06  & \cite{Ia06}  &
  Ia08  & \cite{Ia08}  &
  In77  & \cite{In77}  &
  Is80  & \cite{Is80}  \\
  Ja60  & \cite{Ja60}  &
  Ja78  & \cite{Ja78}  &
  Je07  & \cite{Je07}  &
  Ji02  & \cite{Ji02}  &
  Ka68  & \cite{Ka68}  &
  Ka69  & \cite{Ka69}  \\
  Ke07  & \cite{Ke07}  &\
  Ki89  & \cite{Ki89}  &
  Ki91  & \cite{Ki91}  &
  Ko83  & \cite{Ko83}  &
  Ko87  & \cite{Ko87}  &
  Ko97a & \cite{Ko97a} \\
  Ko97b & \cite{Ko97b} &
  Kr91  & \cite{Kr91}  &
  Le08  & \cite{Le08}  &
  Li94  & \cite{Li94}  &
  Ma94  & \cite{Ma94}  &
  Ma08  & \cite{Ma08}  \\
  Mc67  & \cite{Mc67}  &
  Mi67  & \cite{Mi67}  &
  Mo71  & \cite{Mo71}  &
  Mu04  & \cite{Mu04}  &
  Na91  & \cite{Na91}  &
  Ni69  & \cite{Ni69}  \\
  No74  & \cite{No74}  &
  Oi01  & \cite{Oi01}  &
  Pa72  & \cite{Pa72}  &
  Pa05  & \cite{Pa05}  &
  Pi03  & \cite{Pi03}  &
  Pr67  & \cite{Pr67}  \\
  Pr90  & \cite{Pr90}  &
  Ra83  & \cite{Ra83}  &
  Re85  & \cite{Re85}  &
  Ri07  & \cite{Ri07}  &
  Ro70  & \cite{Ro70}  &
  Ro72  & \cite{Ro72}  \\
  Ro74  & \cite{Ro74}  &
  Ro75  & \cite{Ro75}  &
  Ro06  & \cite{Ro06}  &
  Ry73a & \cite{Ry73a} &
  Ry91  & \cite{Ry91}  &
  Sa80  & \cite{Sa80}  \\
  Sa95  & \cite{Sa95}  &
  Sa04  & \cite{Sa04}  &
  Sa05  & \cite{Sa05}  &
  Sc05  & \cite{Sc05}  &
  Sc07  & \cite{Sc07}  &
  Se73  & \cite{Se73}  \\
  Se05  & \cite{Se05}  &
  Sh55  & \cite{Sh55}  &
  Si66  & \cite{Si66}  & 
  Si72  & \cite{Si72}  & 
  Sq75  & \cite{Sq75}  &
  Sq76  & \cite{Sq76}  \\ 
  Ti95  & \cite{Ti95}  & 
  To03  & \cite{To03}  &
  To05  & \cite{To05}  &
  Vo77  & \cite{Vo77}  &
  Wa83  & \cite{Wa83}  &
  Wa92  & \cite{Wa92}  \\
  We68  & \cite{We68}  & 
  Wh77  & \cite{Wh77}  & 
  Wh81  & \cite{Wh81}  & 
  Wh85  & \cite{Wh85}  &
  Wi76  & \cite{Wi76}  &
  Wi78  & \cite{Wi78}  \\
  Wi80  & \cite{Wi80}  &
  Zi72  & \cite{Zi72}  &
  Zi87  & \cite{Zi87}  &
    
\end{tabular}
\end{ruledtabular}
\end{table*}
\endgroup

For some decays that lead to radioactive daughter nuclei, there is no direct measurement of the
$Q_{EC}$-value for the superallowed transition or the one that exists is rather imprecise.  In these
cases the $Q_{EC}$-value must depend on the measured mass excesses of the parent and daughter nuclei,
together with the excitation energy of the analog 0$^+$ state in the daughter.  Each of these
properties is identified in column 3 of Table~\ref{QEC}, with the individual measurements of that
property, their weighted average and a scale factor appearing in columns to the right.  The average
$Q_{EC}$-value listed for the corresponding superallowed transition is obtained from these separate
averages.  If a direct measurement of the superallowed $Q_{EC}$-value exists, then it is also
included in the final average.

As in our previous survey \cite{HT05}, we have not used the 2003 Mass tables \cite{Au03} to derive
the $Q_{EC}$-values of interest.  Our approach is to include all pertinent measurements for each property;
typically, only a subset of the available data is included as input to the mass tables.  Furthermore, we
have examined each reference in detail and either accepted the result, updated it
to modern calibration standards or rejected it for cause.  The updating procedures are outlined, reference
by reference, in Table~\ref{update} and the rejected results are similarly documented in Table~\ref{reject}.
With a comparatively small data set, we could afford to pay the kind of individual attention that is
impossible when one is considering all nuclear masses.

The half-life data appear in Table~\ref{t1/2} in similar format to Table~\ref{QEC}.  For obvious reasons,
half-life measurements do not lend themselves to being updated.  Consequently, a number of mostly pre-1970
measurements have been rejected because they were not analyzed with the ``maximum-likelihood" method.  The
importance of using this technique for precision measurements was not recognized until that time \cite{Fr69a}
and, without access to the primary data, there is no way a new analysis can be applied retroactively.  All
rejected half-life measurements are also documented in Table~\ref{reject}.

Finally, the branching-ratio measurements are presented in Table~\ref{R}.  The decays of the $T_z = 0$
parents are the most straightforward since, in all these cases, the superallowed branch accounts for
$>$99.5\% of the total decay strength.  Thus, even imprecise measurements of the weak non-superallowed
branches can be subtracted from 100\% to yield the superallowed branching ratio with good precision.
For the higher-Z parents of this type, particularly $^{62}$Ga and heavier, it has been shown theoretically
\cite{Ha02} and experimentally \cite{Pi03, Fi08} that numerous very-weak Gamow-Teller transitions occur,
which, in total, can carry significant decay strength.  Where such unobserved transitions are expected to
exist and have not already been accounted for in the quoted references, we have used a combination of
experiment and theory to account for the unobserved strength, with uncertainties being adjusted accordingly.

The branching ratios for decays from $T_z = -1$ parents are much more challenging to determine, since the
superallowed branch is usually one of several strong branches -- with the notable exception of $^{14}$O --
and, in two of the measured cases, it actually has a branching ratio of less than 10\%.  The decays of
$^{18}$Ne, $^{26}$Si, $^{30}$S, $^{34}$Ar and $^{42}$Ti thus required special treatment.  In each case,
the absolute branching ratio for a single $\beta$-transition has been measured. The branching ratios
for other $\beta$-transitions then had to be determined from the relative intensities of $\beta$-delayed
$\gamma$ rays in the daughter.  The relevant $\gamma$-ray intensity measurements appear in Table~\ref{BDG},
with their averages then being used to determine the superallowed branching-ratio averages shown in
bold type in Table~\ref{R}.  These cases are also flagged with a table footnote.

\section{\label{s:Ft} The $\F t$ Values}

With the input data now settled, we can proceed to derive the $ft$ values for the 20 superallowed
transitions included in the tables.  In our last survey \cite{HT05}, we described and used a new
computer code for calculating the statistical rate function $f$, which surpassed the precision then being
obtained with measurements of $Q_{EC}$.  Since then, with the advent of Penning-trap mass
measurements, experimental uncertainties have shrunk even further.  The level of precision possible
has currently reached $\sim$0.001\%, at least for the $Q_{EC}$ values of $^{50}$Mn and $^{54}$Co, so
it is now necessary to include in the $f$ calculation a provision for atomic excitation of the
daughter nucleus if the calculation is to continue to match the precision of the input data.  Our method
for including this effect is described in Appendix~\ref{s:srf}; and we also present there, in Table~\ref{t:overlap},
a comparison of $f$ values both with and without this small correction.  It can be seen that the effect
of the correction is comparable to a shift of 0.001-0.004\% in the $Q_{EC}$ value, an amount significant
enough to warrant its inclusion in future.  Our final $f$ values are recorded in the second column of
Table~\ref{Ft}.  They were evaluated using our updated code and the $Q_{EC}$ values with their uncertainties
from column 7 of Table~\ref{QEC}.

\begingroup
\squeezetable
\begin{table*}
\begin{center}
\caption{Derived results for superallowed Fermi beta decays.
\label{Ft}}
\vskip 1mm
\begin{ruledtabular}
\begin{tabular}{rrrrrrrr}
& & & & & & & \\[-3mm]
Parent & & $P_{EC}$ &  Partial half-life & &
&  &  \\
nucleus & \multicolumn{1}{c}{$f$} & (\%) &  \multicolumn{1}{c}{$t(ms)$} &  
\multicolumn{1}{c}{$ft(s)$} &
\multicolumn{1}{c}{$\delta_R^{\prime}$ (\%)} & $\delta_C - \delta_{NS}$ (\%) &
 \multicolumn{1}{c}{$\F t (s)$}  \\[1mm] 
\hline
& & & & & & & \\[-3mm]
\multicolumn{2}{l}{$T_z = -1$:} & & & & & & \\
$^{10}$C & $2.3004 \pm 0.0012$ & 0.297 & $1322300 \pm 1800$~ & $3041.7 \pm 4.3$~~~&
$1.679 \pm 0.004$ & $0.520 \pm 0.039$ & $3076.7 \pm 4.6$~~  \\
$^{14}$O & $42.772 \pm 0.023$~$\,$ & 0.088 & $71127 \pm 51$~~~~ & $3042.3 \pm 2.7$~~ &
$1.543 \pm 0.008$ & $0.575 \pm 0.056$ & $3071.5 \pm 3.3$~~ \\
$^{18}$Ne & $134.47 \pm 0.15$~~~ & 0.081 & $21660 \pm 590\,$~~ & $2912 \pm 79$~~~ &
$1.506 \pm 0.012$ & $0.855 \pm 0.052$ & $2931 \pm 80$~~~ \\
$^{22}$Mg & $418.39 \pm 0.17$~~~ & 0.069 & $7295 \pm 17$~~~~ & $3052.0 \pm 7.2$~~ &
$1.466 \pm 0.017$ & $0.605 \pm 0.030$ & $3078.0 \pm 7.4$~~ \\
$^{26}$Si & $1029.4 \pm 2.2$~~~~  & 0.064 & $2954 \pm 23$~~~~ & $3041 \pm 24$~~~ &
$1.438 \pm 0.023$ & $0.650 \pm 0.034$ & $3065 \pm 25$~~~ \\
$^{30}$S  & $1966.9 \pm 8.0$~~~~  & 0.066 & $1524 \pm 21$~~~~ & $2998 \pm 44$~~~ &
$1.423 \pm 0.029$ & $1.040 \pm 0.032$ & $3009 \pm 44$~~~ \\
$^{34}$Ar & $3414.5 \pm 1.5$~~~~  & 0.069 & $894.0 \pm 2.4$~~~ & $3052.7 \pm 8.2$~~ &
$1.412 \pm 0.035$ & $0.845 \pm 0.058$ & $3069.6 \pm 8.5$~~ \\
$^{38}$Ca & $5327.2 \pm 1.8$~~~~ & 0.075 &     &     & 
$1.414 \pm 0.042$ & $0.940 \pm 0.072$ &  \\
$^{42}$Ti & $7040 \pm 30$~~~~~ & 0.088 & $470 \pm 160\,$~~ & $3300 \pm 1100$ &
$1.428 \pm 0.050$ & $1.170 \pm 0.080$ & $3300 \pm 1100$ \\[5mm]
\multicolumn{2}{l}{$T_z = 0 $:} & & & & & & \\
$^{26}$Al$^m$ & $478.237 \pm 0.038\,$ & 0.082 & $6350.2 \pm 1.9$~~~ & $3036.9 \pm 0.9$~~ &
$1.478 \pm 0.020$ & $0.305 \pm 0.027$ & $3072.4 \pm 1.4$~~ \\
$^{34}$Cl & $1995.96 \pm 0.47$~~ & 0.080 & $1527.77 ^{+0.44}_{-0.47}$~~~ & $3049.4 ^{+1.1}_{-1.2}$~~~ & 
$1.443 \pm 0.032$ & $0.735 \pm 0.048$  & $3070.6 \pm 2.1$~~ \\
$^{38}$K$^m$ & $3297.88 \pm 0.34$~~ & 0.085 & $925.42 \pm 0.28\,$~ & $3051.9 \pm 1.0$~~~&
$1.440 \pm 0.039$ & $0.755 \pm 0.060$ & $3072.5 \pm 2.4$~~ \\
$^{42}$Sc & $4472.24 \pm 1.15$~~ & 0.099 & $681.43 \pm 0.26\,$~ & $3047.6 \pm 1.4$~~ &
$1.453 \pm 0.047$ & $0.630 \pm 0.059$ & $3072.4 \pm 2.7$~~ \\
$^{46}$V & $7209.47 \pm 0.90 $~~ & 0.101 & $422.99 \pm 0.11\,$~ & $3049.5 \pm 0.9$~~ &
$1.445 \pm 0.054$ & $0.655 \pm 0.063$ & $3073.3 \pm 2.7$~~ \\
$^{50}$Mn & $10745.97 \pm 0.51$~~ & 0.107 & $283.68 \pm 0.11\,$~ & $3048.4 \pm 1.2$~~ &
$1.444 \pm 0.062$ & $0.695 \pm 0.055$ & $3070.9 \pm 2.8$~~ \\
$^{54}$Co & $15766.6 \pm 2.9\,$~~~ & 0.111 & $193.495 ^{+0.063}_{-0.086}$~~ &$3050.8 ^{+1.1}_{-1.5}~~~$ &
$1.443 \pm 0.071$ & $0.805 \pm 0.068$ & $3069.9 ^{+3.2}_{-3.3}$~~~ \\
$^{62}$Ga & $26400.2 \pm 8.3\,$~~~ & 0.137 & $116.442 \pm 0.042$ & $3074.1 \pm 1.5$~~ & 
$1.459 \pm 0.087$ & $1.52 \pm 0.21$~ & $3071.5 \pm 7.2$~~ \\
$^{66}$As & $32125 \pm 470$~~~ & 0.155 &       &     & 
$1.468 \pm 0.095$ & $1.62 \pm 0.40$~ &       \\
$^{70}$Br & $38600 \pm 3600\,$~ & 0.175 &       &    &
$1.49 \pm 0.11$~ & $1.69 \pm 0.25$~ &       \\
$^{74}$Rb & $47300 \pm 110$~~~ & 0.194 & $65.227 \pm 0.078$ & $3084.9 \pm 7.8$~~ &
$1.50 \pm 0.12$~ & $1.71 \pm 0.31$~ & $3078 \pm 13$~~~ \\[5mm]
& & & & & \multicolumn{2}{r}{Average (best 13), $\overline{\F t}$} & $3072.08 \pm 0.79\,$ \\
& & & & & \multicolumn{2}{r}{$\chi^2/\nu$} & \multicolumn{1}{c}{0.28} \\
\end{tabular}
\end{ruledtabular}
\end{center}
\end{table*}
\endgroup

The third column of Table~\ref{Ft} lists (as percentages) the electron-capture fraction, $P_{EC}$, calculated
for each of the 20 superallowed transitions.  The method of calculation was described in our last survey \cite{HT05},
to which the reader is referred for more details.  The partial half-life, $t$, for each transition is then
obtained from its total half-life, $t_{1/2}$, branching ratio, $R$, and electron-capture fraction according
to the following formula:
\be
t = \frac{t_{1/2}}{R} \left ( 1 + P_{EC} \right ).
\label{partialt}
\ee
The resultant values for the partial half-lives and the corresponding $ft$ values appear in columns 4 and 5
of the table.

To obtain the $\F t$ from each $ft$ value, we use Eq.~(\ref{Ftconst}) to apply the small
transition-dependent correction terms, $\delta_R^{\prime}$, $\delta_{NS}$ and $\delta_C$.  We take the values
of these terms from our recent re-evaluation of the corrections to superallowed beta decay \cite{TH08}.
The first term, $\delta_R^{\prime}$, which is listed in column 6 of Table~\ref{Ft}, is taken from Table V
in Ref.~\cite{TH08}.  The two nuclear structure-dependent terms, combined in the form ($\delta_C -
\delta_{NS}$), are listed in column 7.  In Ref.~\cite{TH08}, $\delta_C$ is expressed as the sum of
$\delta_{C1}$ and $\delta_{C2}$, the former being listed in Table III of that reference and the latter in
Table II; $\delta_{NS}$ is taken from Table VI of the same reference.  Finally, the resulting $\F t$ values
are listed in column 8 of our Table~\ref{Ft}.

Both the uncorrected $ft$ values and the fully corrected $\F t$ values are plotted in Fig.~\ref{ftFt}
for the 13 most precisely measured transitions.  The differences between the former, in the top panel, and
the latter, in the bottom panel, illustrate the effects of our including the correction terms.  It is also
worth remarking that the values of $\delta_R^{\prime}$ are very nearly the same for 11 of the 13 cases
plotted: only $^{10}$C and $^{14}$O have slightly higher values.  Thus, most of the differences between
the two panels of the figure are due to the effects of the nuclear structure-dependent terms, $\delta_{NS}$
and $\delta_C$. 

\subsection{CVC test}
\label{ss:cv}

\begin{figure}[t]
\epsfig{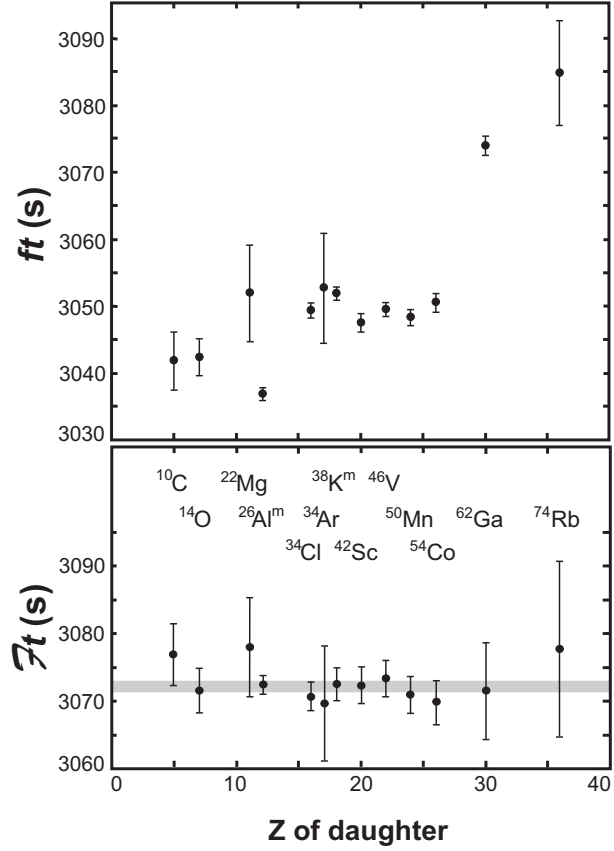}
\caption{In the top panel are plotted the uncorrected experimental $ft$ values as a function of the
charge on the daughter nucleus.  In the bottom panel, the corresponding $\F t$ values are given; they
differ from the $ft$ values by the inclusion of the correction terms $\delta_R^{\prime}$, $\delta_{NS}$
and $\delta_C$.  The horizontal grey band in the bottom panel gives one standard deviation around the
average $\overline{\F t}$ value.}
\label{ftFt}
\end{figure}

There are now 13 superallowed transitions whose $\F t$ values have uncertainties less than
$\pm$0.4\%, with the best case, $^{26}$Al$^m$, being known an order of magnitude better than that.
These data are sufficient to provide a very demanding test of the CVC assertion that the $\F t$ values should
be constant for all nuclear superallowed transitions of this type.  The data in column 8 of Table~\ref{Ft} clearly
satisfy the test, the weighted average of the 13 most precise results being
\be
\overline{\F t} = 3072.08 \pm 0.79 s,
\label{Ftavgstat}
\ee
with a corresponding chi-square per degree of freedom of $\chi^2/\nu = 0.28$.  That these 13 $\F t$ values
form a consistent set is also clearly evident from the bottom panel of Fig.~\ref{ftFt}.  Since $\F t$
is proportional to the {\it square} of the vector coupling constant, $\GV$, then Eq.~(\ref{Ftavgstat}) can be
said to confirm the constancy of $\GV$ -- and to verify this key component of the CVC hypothesis -- at the level
of $1.3\times10^{-4}$.

Compared with the results of our last survey \cite{HT05}, the value of $\overline{\F t}$ in Eq.~(\ref{Ftavgstat})
is somewhat lower but carries a similar uncertainty.  However, the new analysis is more demanding since, for
the first time, it includes the $^{62}$Ga transition, which has been very significantly improved in the last
four years.  This effectively increases the span of masses over which the CVC test is being applied; yet even
with this addition, the $\chi^2/\nu$ is actually lower than it was previously.  The small reduction in the
central value of $\overline{\F t}$ is within the uncertainty of the previous value; it has arisen principally
from recent changes in the nuclear structure-dependent correction terms \cite{TH08}.

\begin{figure}[t]
\epsfig{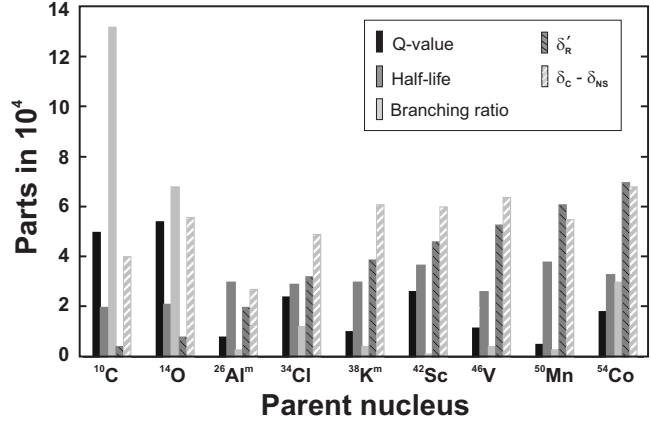}
\caption{Summary histogram of the fractional uncertainties attributable to each 
experimental and theoretical input factor that contributes to the final
$\protect\F t$ values for the ``traditional nine" superallowed transitions.}
\label{f:hist9}
\end{figure}

\begin{figure}[b]
\epsfig{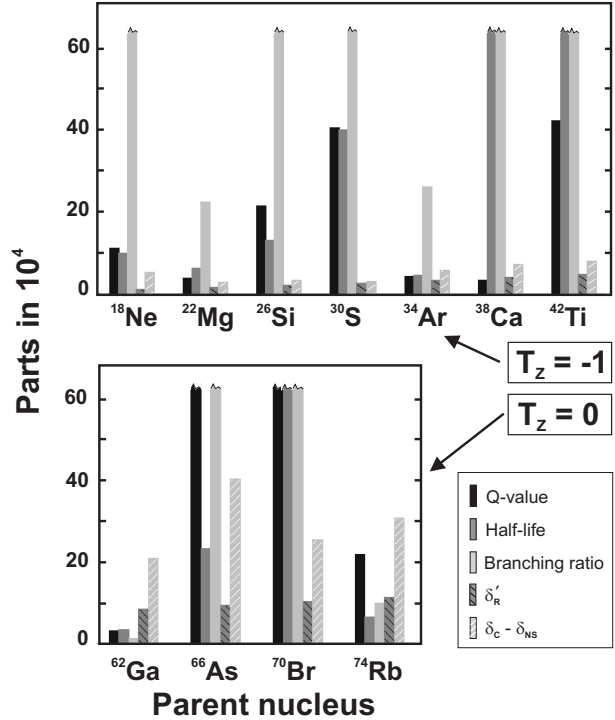}
\caption{Summary histogram of the fractional uncertainties attributable to each 
experimental and theoretical input factor that contributes to the final
$\protect\F t$ values for the eleven other superallowed transitions.  Where
the error is shown as exceeding 60 parts in $10^4$, no useful experimental
measurement has been made.}
\label{f:hist12}
\end{figure}

\subsection{$\F t$ value error budgets}
\label{ss:su}

We show the contributing factors to the individual $\F t$-value uncertainties
in Fig.~\ref{f:hist9} for the "traditional nine" cases and in Fig.~\ref{f:hist12}
for the remaining eleven.  For most of the cases that contribute to the CVC test --
$^{26}$Al$^m$ to $^{54}$Co in Fig.~\ref{f:hist9} as well as $^{62}$Ga and $^{74}$Rb in
Fig.~\ref{f:hist12} -- the theoretical uncertainties are greater than, or comparable to,
the experimental ones.  In these cases, the nuclear-structure-dependent correction,
$\delta_C - \delta_{NS}$, contributes an uncertainty of 3-7 parts in $10^4$ to all
$\F t$ values between $^{26}$Al$^m$ and $^{54}$Co but jumps up to 20-30 parts in
$10^4$ for $^{62}$Ga and $^{74}$Rb because of nuclear-model ambiguities.  For its part,
the nucleus-dependent radiative correction, $\delta_R^{\prime}$, has an
uncertainty that starts very small but grows smoothly with $Z^2$.  This is because the
contribution to $\delta_R^{\prime}$ from order $Z^2 \alpha^3$ has only been estimated from
its leading logarithm \cite{Si87} and the magnitude of this estimate has been taken as the
uncertainty in $\delta_R^{\prime}$.  As a result, though, for $^{50}$Mn and $^{54}$Co it
becomes the leading uncertainty, indicating that a closer look at the order $Z^2 \alpha^3$
contribution to $\delta_R^{\prime}$ would certainly now be worthwhile.

For all the transitions from $T_z$=0 parent nuclei, the experimental branching ratios
are $> 99 \%$ and have very small associated uncertainties with the exception
of $^{54}$Co, which has a $3\times10^{-4}$ fractional uncertainty attributed to its branching
ratio, and $^{74}$Rb, which has $10\times10^{-4}$.  In both cases, this is because they are
predicted to have weak Gamow-Teller branches that have not yet been observed.  We have
used an estimate of the strength of the missing branches, taken from a shell-model
calculation \cite{Ha02}, to assign an uncertainty to the superallowed branching ratio.
Numerous weak Gamow-Teller branches become an increasingly significant issue for the
heavier-mass nuclei with $A \geq 62$, where they present a major experimental challenge
if they are to be fully characterized.  Only in the case of $^{62}$Ga has this been
accomplished so far.  

For the decays of $^{10}$C and $^{14}$O, and for all the decays depicted in Fig.~\ref{f:hist12}
except for $^{62}$Ga and $^{74}$Rb, the predominant uncertainties are experimental in origin.
Many of the experimental branching ratios, and some of the $Q$-values and half-lives have
yet to be measured precisely for the cases in Fig.~\ref{f:hist12}, but recent advances in experimental
techniques have been improving this situation and are likely to improve it even more within
the next few years.

\subsection{Accounting for systematic uncertainties}
\label{ss:se}

So far, we have dealt with the inter-nuclear behavior of $\F t$-values,
examining their constancy as a test of CVC.  With that test passed
at high precision, we are now in a position to use the average
$\F t$-value obtained from these concordant nuclear data to go
beyond nuclei, obtaining first the vector coupling constant (see
Eq.~(\ref{Ftconst})) and then the $V_{ud}$ matrix element.  Before
doing so, however, we must address one more possible source of
uncertainty.  Though the average $\F t$ value given in Eq.~(\ref{Ftavgstat})
includes a full assessment of the uncertainties attributable to
experiment and to the particular calculations used to obtain the
correction terms, it does not incorporate any provision for a
common systematic error that could arise from the type of
calculation chosen to model the nuclear-structure effects.  
To discuss this, we divide the problem into two parts:  the accuracy
of the model as an approximation to the formally complete treatment;
and the possible existance of systematic uncertainties within the model.

\subsubsection{The model approximation}
\label{sss:fca}

Very recently Miller and Schwenk \cite{MS08} have explored the formally complete
approach to isospin-symmetry breaking.  Their starting point is to define the Fermi matrix
element as
\be
M_F = \langle f | \tau_+ | i \rangle =
\sum_{\alpha} \langle f | a_{\alpha}^{\dag} b_{\alpha} | i \rangle =
\sum_{\alpha , \pi} \langle f | a_{\alpha}^{\dag} | \pi \rangle
\langle \pi | b_{\alpha} | i \rangle ,
\label{MF}
\ee
where $a_{\alpha}^{\dag}$ creates a neutron and $b_{\alpha}$ annihilates a proton in state
$\alpha$.  Here $| i \rangle$ and $| f \rangle$ are the {\em exact} state vectors for the
full Hamiltonian.  If this Hamiltonian commutes with the isospin operators, then $|i \rangle$
and $| f \rangle$ are exact isospin analogs of each other, $\langle \pi | b_{\alpha} | i \rangle =
\langle f | a_{\alpha}^{\dag} | \pi \rangle^{\ast}$ and the symmetry-limit matrix element is
\be
M_0 = \sum_{\alpha , \pi} | \langle f | a_{\alpha}^{\dag} | \pi \rangle |^2 .
\label{M0}
\ee
If isospin is not an exact symmetry, then $|i \rangle$ and $| f \rangle$ are not isospin analogs
and a correction to $M_0$ needs to be evaluated.  This is the isospin-symmetry-breaking correction,
$\delta_C$, we seek to determine.  It is defined by
\be
M_F^2 = M_0^2 \left ( 1 - \delta_C \right ) .
\label{MFC}
\ee
Ideally, to obtain $\delta_C$ one would compute Eq.~(\ref{MF}) using the shell model, and introduce
Coulomb and other charge-dependent terms into the shell-model Hamiltonian.  However, because the
Coulomb force is long range, the shell-model space would have to be huge to include all the potential
states that the Coulomb interaction might connect with.  Currently this is not a practical proposition.

To proceed with a manageable calculation, we have developed a model approach \cite{To77,TH02,TH08}
in which $\delta_C$ is divided into two parts:
\be
\delta_C = \delta_{C1} + \delta_{C2} .
\label{dc12}
\ee
For $\delta_{C1}$, we compute
\be
\sum_{\alpha , \pi} \langle \overline{f} | a_{\alpha}^{\dag} | \pi
\rangle \langle \pi | b_{\alpha} | \overline{\imath} \rangle
= M_0 \left ( 1 - \delta_{C1} \right )^{1/2}  ,
\label{MF1}
\ee
where $| \overline{\imath} \rangle$ and $| \overline{f} \rangle$ are not the exact eigenstates that appear
in Eq.~(\ref{MF}), but are the shell-model eigenstates of an effective  Hamiltonian (including
charge-dependent terms) evaluated in a modest-sized shell-model space.  Since this space does not
allow for nodal mixing, we correct for that limitation by computing the second component,
$\delta_{C2}$, obtained from
\be
\sum_{\alpha , \pi} | \langle \overline{f} | a_{\alpha}^{\dag} |
\pi \rangle |^2 r_{\alpha}^{\pi} 
= M_0 \left ( 1 - \delta_{C2} \right )^{1/2} ,
\label{MF2}
\ee
where each $r_{\alpha}^{\pi}$ is a radial overlap integral of proton and neutron radial functions.  
We justify the efficacy of this second term by the following arguments:  If the radial functions
were identical, then $\delta_{C2}$ would vanish as it should.  Otherwise the proton radial functions, 
$u^p(r)$, could be expanded in terms of a complete set of neutron functions, $u_N^n(r)$, including
all possible radial nodes, $N$:
\be 
u^p(r) = \sum_N a_N u_N^n(r) .
\label{upunN}
\ee
The isospin-symmetry breaking correction, $\delta_{C2}$, could then be expressed in terms of
$a_N$, which from perturbation theory could itself be written in terms of matrix elements of the
Coulomb interaction.  This would be equivalent to the nodal mixing included in Eq.~(\ref{MF}) but
left out of the calculation of $\delta_{C1}$ in Eq.~(\ref{MF1}).  The idea is that $\delta_{C1}$ is
the result of a tractable shell-model calculation that does not include any nodal mixing, while
$\delta_{C2}$ then corrects for the nodal mixing that would be present if the shell-model space were
larger.

Clearly our charge-dependent correction terms \cite{To77,TH02,TH08} are based on a model, and
required approximations to make the computation possible.  Since no one has yet made a complete
calculation without approximations, it is impossible to be definitive about any systematic errors
that might be introduced by our methods.  Only for the lightest superallowed emitter, $^{10}$C,
has it been possible so far even to come close to an exact treatment.  Caurier {\it et al.} \cite{Ca02}
have reported a large no-core shell-model calculation for that system but, even though they
were able to extend their basis states up to $8\hbar\omega$, their calculated $\delta_C$ still had
not converged to a stable value.  However they used their results together with perturbation theory
to estimate that the full value of $\delta_C$ should be about 0.19\%.  This result, which in effect
used Eq.~(\ref{MF}) and did not split $\delta_C$ into two parts, agrees completely with our calculated
value for $\delta_C=\delta_{C1}+\delta_{C2}$ of 0.18(2)\% (see Table VII in Ref.~\cite{TH08}).  
This agreement certainly supports the validity of our model.

Furthermore, it must be noted that our model approach has allowed us to use well-established shell-model
and related parameters, which were determined from experimental data that are completely independent
of the superallowed $ft$ values.  As is clearly evident from Fig.~\ref{ftFt}, these calculated
corrections do a remarkable job in converting widely scattered $ft$ values into a consistent set of
$\F t$ values.  Not only that but, as shown in Ref.~\cite{TH08}, they also closely reproduce the measured
results for isospin-forbidden $0^+$$\rightarrow 0^+$ $\beta$ transitions in all nuclei for which the
shell-model calculation is well specified.  (This is not the case for $^{62}$Ga.)  Of
course, although these successes demonstrate that our calculated $\delta_C$ values correctly reproduce
the nucleus-to-nucleus variations observed by experiment, they cannot rule out a constant shift in
the corrections for all nuclei.  Even so, it seems highly unlikely that a faulty approximation could
lead to relative results that are correct in every detail, while being consistently wrong -- and by the
same constant amount -- in the absolute values for each and every case.  

Under the circumstances, we see no justification at this time to assign any additional systematic error
to account for possible inadequacies of the model we use to calculate the charge-dependent correction terms.

\subsubsection{Systematic uncertainty within the model}
\label{sss:sysinm}

As introduced in Eq.~(\ref{dc12}) the isospin-symmetry breaking correction, $\delta_C$, is
separated into two pieces:  $\delta_{C1}$ comes from configuration mixing in a modest-sized
shell-model calculation with charge-dependent interactions, while $\delta_{C2}$ involves radial
overlap integrals.  The calculation of $\delta_{C1}$, the smaller of the two terms, requires
a reliable shell-model description of the nuclei involved but it can also be further constrained
by independent experimental data: for example, the measured coefficients of the corresponding
isobaric multiplet mass equation (IMME).  We take our corrections from Ref.~\cite{TH08} where
the uncertainties attached to the calculated values of $\delta_{C1}$ for the twenty cases of
interest here already include ample provision for differences between competing shell-model
parameterizations as well as for experimental uncertainties on the IMME coefficients used as
constraints.

The values of the radial-overlap term, $\delta_{C2}$, which we use as input to Table~\ref{Ft},
were also taken from our recently published calculations \cite{TH08}.  Those calculations used
radial wave functions derived from a Saxon-Woods potential with either the well depth or one of
the surface terms in the potential adjusted so that the binding energy of each computed
eigenfunction matched the corresponding measured separation energy.  The quoted uncertainties
included provision for any variations in the results depending on which parameter was used in
the adjustment.  However no provision was included for possible differences that might occur
if another method entirely were used to derive the radial wave functions.  In the past \cite{HT05},
we have accounted for this uncertainty by comparing our results for $\delta_{C2}$ with those of
Ormand and Brown \cite{OB85,OB89,OB95}, who used Hartree-Fock eigenfunctions and obtained consistently
smaller corrections than those we found with a Saxon-Woods potential.  We treated this as a valid
source of systematic uncertainty and incorporated it by deriving two average $\overline{\F t}$ values,
one for each set of $\delta_{C2}$ calculations, then taking the average of the two and assigning a
systematic uncertainty equal to half the spread between them.

This specific comparison is no longer tenable.  The Ormand and Brown calculations are in some cases
more than two decades old: they use smaller shell-model spaces than are now known to be necessary
\cite{TH08} and they are not available at all for some of the transitions that we now need to include.
To remedy these deficiencies we have undertaken our own Hartree-Fock calculations.  They are described
in detail in Appendix~\ref{s:syserror}, where Table~\ref{ss:newHF} lists the values of $\delta_{C2}$
we compute from Hartree-Fock-derived wave functions and compares them with our earlier results
from the Saxon-Woods potential \cite{TH08}, the same results that we used to evaluate $\overline{\F t}$
in Table~\ref{Ft}.  Both methods used exactly the same shell-model calculations to determine the full
parentage of the states involved.

With these new Hartree-Fock calculations we can now follow a similar procedure to the one we employed with
the old calculations in our previous survey \cite{HT05}.  We begin by substituting the Hartree-Fock
$\delta_{C2}$ values for the Saxon-Woods ones in deriving the $\delta_C$ values used in Table~\ref{Ft}.
When we do this the $\overline{\F t}$-value result becomes 3071.60$\pm$0.89 with $\chi^2/\nu$=0.98.  This
normalized chi-square is three times worse than the one we obtained in Table~\ref{Ft} with the Saxon-Woods
corrections, which arguably could justify our rejecting the Hartree-Fock results outright.  However, to be
safe, we proceed as before and take the average of the Hartree-Fock and Saxon-Woods results, adding a
systematic uncertainty equal to half the spread between the two results.  Thus, we obtain
\bea
\overline{\F t} & = & 3071.87 \pm 0.79_{\rm stat} \pm 0.27_{\rm syst}~{\rm s}
\nonumber \\
& = & 3071.87 \pm 0.83~{\rm s} ,
\label{Ftavg}
\eea
where on the second line the two uncertainties have been added in quadrature.  Our new systematic adjustment
amounts to only 0.27s, much smaller and of opposite sign to the 0.90s correction applied previously
\cite{HT05}.

It is the value for $\overline{\F t}$ in Eq.~(\ref{Ftavg}) that we carry forward to subsequent sections
where we obtain $V_{ud}$ and test the unitarity of the CKM matrix.

\section{\label{s:impact} The Impact on Weak-Interaction Physics}

\subsection{The Value of $V_{ud}$}
\label{ss:vVud}

With a mutually consistent set of $\F t$ values, we can now use the adjusted average value in
Eq.~(\ref{Ftavg}) to determine the vector coupling constant, $\GV$, from Eq.~(\ref{Ftconst}).  
The value of $\GV$ itself is of little interest but, together with the weak interaction constant
for the purely leptonic muon decay, $\GF$, it yields the much more interesting up-down element
of the Cabibbo-Kobayashi-Maskawa (CKM) quark-mixing matrix.  The basic relationship is $V_{ud}=
\GV/\GF$, which in terms of $\overline{\F t}$ becomes
\be
V_{ud}^2 = \frac{K}{2 \GF^2 (1 + \DRV ) \overline{\F t}},
\label{Vudeq}
\ee
where $\DRV$ is the nucleus-independent radiative correction.  This correction has recently
been carefully re-examined by Marciano and Sirlin \cite{Ma06}, who very substantially reduced
its uncertainty.  Expressing their new result in a way that is consistent with the definition
of our other correction terms, we obtain (see Eq.~(41) in Ref.~\cite{TH08}) 
\be
\DRV = (2.361 \pm 0.038 ) \%.
\label{DRV}
\ee

Using the Particle Data Group (PDG) \cite{PDG08} value for the weak interaction coupling
constant from muon decay of $\GF /(\hbar c )^3 = (1.16637 \pm 0.00001) \times 10^{-5}$
GeV$^{-2}$, we obtain from Eq.~(\ref{Vudeq}) the result
\be
|V_{ud}|^2 = 0.94916 \pm 0.00044.
\label{Vudsq}
\ee
Note that the total uncertainty here is almost entirely due to the uncertainties
contributed by the theoretical corrections.  By far the largest contribution, 0.00035,
arises from the uncertainty in $\DRV$; 0.00020 comes from the nuclear-structure-dependent
corrections $\delta_C - \delta_{NS}$ and 0.00004 is attributable to $\delta_R^{\prime}$.
Only 0.00016 can be considered to be experimental in origin.

\begin{figure}
\epsfig{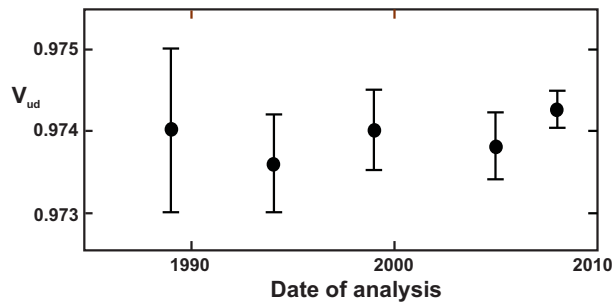}
\caption{Values of $V_{ud}$ as determined from superallowed $0^+$$\rightarrow 0^+$ $\beta$ decays
plotted as a function of analysis date, spanning the past two decades.  In order, from the earliest
date to the most recent, the values are taken from Refs.~\cite{HT90}, \cite{TH95}, \cite{TH99},
\cite{HT05} and this work.}
\label{f:vudvalu}
\end{figure}

The corresponding value of $V_{ud}$ is
\be
|V_{ud}| = 0.97425 \pm 0.00022,
\label{Vudvalu}
\ee
a result which is consistent with, but more precise than, values we have obtained in previous
analyses of superallowed $\beta$ decay.  To emphasize the consistency and steady improvement
that has characterized the value of $V_{ud}$ as derived from nuclear $\beta$ decay, in
Fig.~\ref{f:vudvalu} we plot our new result together with $V_{ud}$ values published at various
times over the past two decades \cite{HT90,TH95,TH99,HT05}.  Clearly the large number of
measurements that contribute to the nuclear determination of $V_{ud}$ provides a robust data
base, one not subject to sudden shifts as each new measurement appears.

\subsection{Unitarity of the CKM matrix}
\label{ss:unitarity}

The Cabibbo-Kobayashi-Maskawa (CKM) matrix transforms one set of quark basis states into another:
it transforms the quark-mass eigenstates into the weak-interaction eigenstates.  If both sets are
complete and orthonormal, then the transformation matrix itself must be unitary.  The Standard Model
does not prescribe the individual elements of the CKM matrix -- they must be determined
experimentally -- but absolutely fundamental to the model is the requirement that the matrix be
unitary.  Whether unitarity is satified in practice can be tested experimentally, the severity of
the test depending on the precision with which the CKM matrix elements can be determined.

To date, the most demanding test of CKM unitarity comes from the sum of squares of the top-row
elements, $|V_{ud}|^2 + |V_{us}|^2 + |V_{ub}|^2$, which should equal one.  Since $|V_{ud}|^2$
constitutes $95 \%$ of this sum, the precision on $V_{ud}$ is of paramount importance.  The
value of $|V_{ud}| = 0.97425(22)$ derived in Sect.~\ref{ss:vVud} has a precision of $0.02 \%$
which is the most precise result so far obtained for this matrix element and is, by more than
an order of magnitude, the most precisely determined value for any element in the CKM matrix.
Alternative methods of obtaining $V_{ud}$ from neutron beta decay, $V_{ud} = 0.9746(18)$, and
from pion beta decay, $V_{ud} = 0.9749(26)$ -- both values taken from the Particle Data
Group's 2008 compilation \cite{PDG08} -- are much less precise and have been hampered by experimental
difficulties.  In the case of the neutron, not only is physical containment a problem but the
axial-vector contribution to its $\beta$ decay must be separated from the vector contribution
by a beta-asymmetry measurement; while for pion beta decay a very small branching fraction,
${\cal O}(10^{-8})$, must be measured.  

At the time of our last survey \cite{HT05} the value of $V_{us}$ was in a state of flux.  The
2004 Particle Data Group value was $|V_{us}| = 0.2200(26)$, based mostly on measurements that
were at least two decades old, but new results then emerging were suggesting a value some
two standard deviations higher.  In the last four years, these new results on the semi-leptonic
decays, $K_{\ell 3}$, of both charged and neutral kaons -- from BNL-E865 \cite{Sh03}, KTeV
\cite{Al04}, NA48 \cite{La04}, KLOE \cite{Am06}, and ISTRA+ \cite{Ro07} -- have all combined
to clarify the situation.  Now, current averages by the 2008 Particle Data Group \cite{PDG08} and
FlaviaNet \cite{Fl08} for kaon semi-leptonic branching fractions are based only on recent, 
high-statistics experiments, which are also consistent with one another.  The best current
value, presented at the CKM2008 Workshop \cite{CKM08} by the FlaviaNet group, is
\be
f_+(0) |V_{us}| = 0.21673 \pm 0.00046 .
\label{fVus}
\ee 
Here $f_+(0)$ is the semi-leptonic decay form factor at zero-momentum transfer.  Its value
is close to unity.  In fact, the CVC hypothesis in the exact SU(3) symmetry limit establishes
its value to be exactly one, but SU(3) symmetry is broken to some extent and a theoretical
calculation is required to estimate the departure of $f_+(0)$ from unity.  Currently, there
are two classes of evaluation: analytic or semi-analytic approaches based on chiral perturbation
theory \cite{LR84,BT03,Ja04,Ci05,Po07}, and those based on lattice QCD \cite{Bo07,Ju07,Ka07}.
We will follow the FlaviaNet group and adopt the lattice value of $f_+(0) = 0.9644 \pm 0.0049$
from the RBC-UKQCD collaboration \cite{Bo07}, which yields
\be
|V_{us}| = 0.2247 \pm 0.0012 .
\label{VusKl3}
\ee

An independent determination of $V_{us}$ can be obtained from the purely leptonic decay of the
kaon, the most important mode being $K^+ \rightarrow \mu^+ \nu$.  If it is considered as a ratio
with the leptonic decay of the pion, $\pi^+ \rightarrow \mu^+ \nu$, the hadronic uncertainties
can be minimized and the result yields the ratio of the CKM matrix elements $|V_{us}|/|V_{ud}|$.
In the analysis of the FlaviaNet group \cite{Fl08} the current result is
\be
\frac{|V_{us}|}{|V_{ud}|} \times \frac{f_K}{f_{\pi}} = 0.2760 \pm 0.0006 ,
\label{fVusVud}
\ee
where $f_K$ and $f_{\pi}$ are the kaon and pion decay constants.  This ratio of pseudoscalar
decay constants has to be obtained from theory, for which lattice QCD seems to be the only reliable
source.  Again, following the FlaviaNet group we adopt the lattice result from the MILC-HPQCD
collaboration \cite{Fo08}, $f_K/f_{\pi} = 1.189 \pm 0.007$, and obtain
\be
\frac{|V_{us}|}{|V_{ud}|} = 0.2321 \pm 0.0005 .
\label{VusVud}
\ee

Thus, we now have three pieces of data -- $|V_{ud}|$ from nuclear decays, Eq.~(\ref{Vudvalu}),
$|V_{us}|$ from $K_{\ell 3}$ decays, Eq.~(\ref{VusKl3}), and the ratio $|V_{us}|/|V_{ud}|$ from
$K_{\ell 2}$ decays, Eq.~(\ref{VusVud}) -- from which to determine two parameters, $|V_{ud}|$
and $|V_{us}|$.  We perform a non-linear least squares fit to obtain the result
\be
|V_{ud}| = 0.97424(22) ~~~~~ |V_{us}| = 0.22534(93) .
\label{fit}
\ee
Note that the value of $|V_{ud}|$ obtained from this fitting procedure has only changed by one
unit in the last figure compared to Eq.~(\ref{Vudvalu}); and the change in $|V_{us}|$ compared to
Eq.~(\ref{VusKl3}), though somewhat larger, is still well within the quoted uncertainties.

The third element of the top row of the CKM matrix, $V_{ub}$, is very small and hardly impacts
on the unitarity test at all.  Its value from the 2008 Particle Data Group compilation \cite{PDG08}
is $|V_{ub}| = (3.93 \pm 0.35) \times 10^{-3}$.  Combining this number with the ones in
Eq.~(\ref{fit}) we find the sum of the squares of the top-row elements of the CKM matrix to be
\be
|V_{ud}|^2 + |V_{us}|^2 + |V_{ub}|^2 = 0.99995 \pm 0.00061 ,
\label{usum}
\ee
a result that shows unitarity to be fully satisfied at the $0.06 \%$ level.  Only $V_{us}$ and
$V_{ud}$ contribute perceptibly to the uncertainty and their contributions are almost equal to one
another.  This may seem surprising since $V_{ud}$ is known to much higher precision than $V_{us}$,
but it follows from the fact that $|V_{ud}|^2$ contributes 95\% to the unitarity sum.

\subsection{Limit on Scalar Interactions}
\label{ss:si} 

\subsubsection{Fundamental scalar current}
\label{sss:fsc}

In our previous survey \cite{HT05} we explained in detail how a scalar current, if it existed,
would affect the $\F t$-value data.  We demonstrated that its effect on $\F t$ would be
approximately proportional to $\langle1/W\rangle$, the average inverse
decay energy of each $\beta^+$ transition, so its presence would be manifest by $\F t$ values
that are not constant as a function of $Z$.  Since $\langle1/W\rangle$ increases monitonically
as $Z$ decreases, the largest deviation of $\F t$ from constancy would occur for the superallowed
transitions from nuclei with the lowest $Z$, $^{10}$C and $^{14}$O.

\begin{figure}[t]
\epsfig{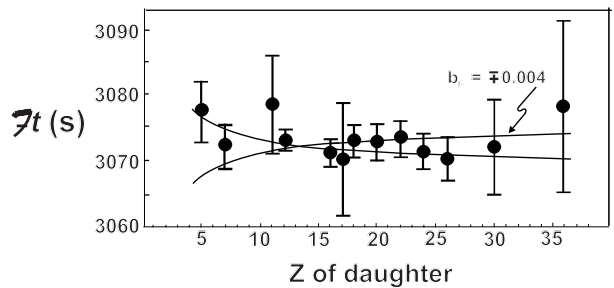}
\caption{Corrected $\F t$ values from Table~\ref{Ft} plotted as a function of the charge on the
daughter nucleus, $Z$.  The curved lines represent the approximate loci the $\F t$ values would
follow if a scalar current existed with $b_F = \pm 0.004$.}
\label{f:scalar}
\end{figure}

We have now repeated the same analysis on our new survey results.  We evaluated the statistical
rate functions, $f$, with a shape-correction factor that included the presence of a scalar
current via the Fierz interference term, $b_F$, which we treated as an adjustable parameter.  We
then sought the value of $b_F$ that minimized $\chi^2$ in a least-squares fit to the expression
$\F t$ = constant.  The result we obtained is
\be
b_F = -0.0022 \pm 0.0026 , 
\label{bF}
\ee
which is consistent with zero, as it was in 2005 \cite{HT05}.  In Fig.~\ref{f:scalar} we illustrate
the sensitivity of this analysis by plotting the measured $\F t$ values together with the loci of
$\F t$ values that would be expected if $b_F = \pm 0.004$.  Obviously, the measured $\F t$ values
do not exhibit any statistically significant curvature.

The result in Eq.~(\ref{bF}) can also be expressed in terms of the coupling constants that
Jackson, Treiman and Wyld \cite{JTW57} used in writing a general form for the weak-interaction
Hamiltonian.  Since we are dealing only with Fermi superallowed transitions, we can restrict
ourselves to scalar and vector couplings, for which that Hamiltonian becomes the following:
\bea
H_{S+V} &  =  & (\overline{\psi}_p \psi_n)
        (C_S \overline{\phi}_e \phi_{\overline{\nu}_e}
        + C_S^{\prime} \overline{\phi}_e \gamma_5 \phi_{\overline{\nu}_e})
\nonumber \\
& &
+ \left ( \overline{\psi}_p
\gamma_{\mu}  \psi_n \right )
\left [C_V \overline{\phi}_e \gamma_{\mu} 
( 1 + \gamma_5 ) \phi_{\overline{\nu}_e} \right ]  ,
\label{JTWSV}
\eea
where we have taken the vector current to be maximally parity violating, as indicated by
experiment \cite{Se06}.  The complexity of the relationship between $b_F$ and the couplings
$C_S$, $C_S^{\prime}$ and $C_V$ depends on what assumptions are made about the properties
of the scalar current.  If we take the most restrictive conditions, that the scalar and
vector currents are time-reversal invariant ({\it i.e.} $C_S$ and $C_V$ are real) and that
the scalar current, like the vector current, is maximally parity violating ({\it i.e.}
$C_S = C_S^{\prime}$), then we can write
\be
\frac{C_S}{C_V} = -\frac{b_F}{2} = +0.0011 \pm 0.0013 .
\label{CS/CV}
\ee
This limit from superallowed $\beta$ decay is, by far, the tightest limit on the presence of a
scalar current under the assumptions stated.

\begin{figure}
\epsfig{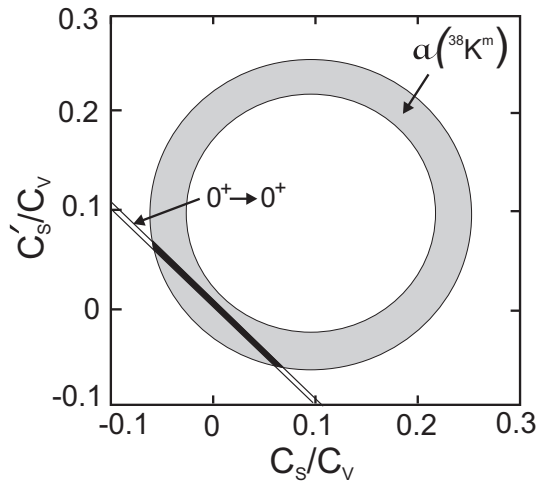}
\caption{Allowed range of values for $C_S/C_V$ and $C_S^{\prime}/C_V$ as determined from the $^{38}$K
$\beta$-$\nu$ angular-correlation measurement \cite{Go05} (grey annulus) and from superallowed
$0^+$$\rightarrow 0^+$ $\beta$ decays (narrow diagonal band).  The overlap region is shaded in
black.  We plot the 68\% confidence limits.}
\label{f:CSCV}
\end{figure}

If we remove the condition that the scalar current be maximally parity violating, then the
expression contains two unknowns,
\bea
b_F & = & \frac{-2 C_V(C_S + C_S^{\prime})}{2|C_V|^2 + |C_S|^2 + |C_S^{\prime}|^2}
\nonumber \\
& \simeq & - \left (\frac{C_S}{C_V} + \frac{C_S^{\prime}}{C_V} \right ) ,
\label{bFparity}
\eea 
and cannot be solved individually for $C_S/C_V$ and $C_S^{\prime}/C_V$.  However, the $\beta$-$\nu$
angular-correlation coefficient, $a$, for a superallowed $0^+$$\rightarrow 0^+$ $\beta$ transition
provides another independent measure of $C_S$ and $C_V$.  In that case,
\bea
a & = & \frac{2|C_V|^2 - |C_S|^2 - |C_S^{\prime}|^2)}{2|C_V|^2 + |C_S|^2 + |C_S^{\prime}|^2}
\nonumber \\
& \simeq & 1 - \frac{1}{2} \left ( \left |\frac{C_S}{C_V} \right |^2
+ \left | \frac{C_S^{\prime}}{C_V} \right |^2 \right ) ,
\label{aparity}
\eea  
which, together with Eq.~(\ref{bFparity}), can be used set limits on both $C_S/C_V$ and
$C_S^{\prime}/C_V$.  Currently, the most precise measurement of such a $\beta$-$\nu$ angular
correlation is for the superallowed decay of $^{38}$K \cite{Go05}.  In this case, what was
actually measured is {\it \~a} = $a/(1 + \gamma b_F m_e/\langle W\rangle)$, where
$\gamma = \sqrt{1-(\alpha Z)^2}$ and $m_e$ is the mass of the electron\footnote{Our
$b_F$ is defined differently from the $b$ used in Ref.~\cite{Go05}.  The two are related
by $b_F = b/\gamma$.}.  The results in
terms of $C_S/C_V$ and $C_S^{\prime}/C_V$ are plotted in Fig.~\ref{f:CSCV}.  The value of
{\it \~a} taken from Ref.~\cite{Go05} leads to the grey annulus plotted in the figure, while
our result for $b_F$ from Eq.~(\ref{bF}) is responsible for the narrow diagonal band.  The
intersection of these two regions, which is in black, defines the 68\% confidence limit
(one standard deviation) for $C_S/C_V$ and $C_S^{\prime}/C_V$.  It corresponds to the limit
\be
\left |\frac{C_S}{C_V} \right | \leq 0.065 ,
\label{CSlimit}
\ee
and exactly the same limit for $C_S^{\prime}/C_V$.

The Jackson-Treiman-Wyld Hamiltonian \cite{JTW57} is a parameterization of one possible extension
to the Standard Model.  The coupling constants $C_S$ and $C_S^{\prime}$ are not prescribed by the
authors but are simply parameters that must be determined from experiment.  One model that actually
introduces scalar interactions in a natural way is the minimal supersymmetric standard model (MSSM).
A valuable review of low-energy tests of this model has recently been published by Ramsey-Musolf
and Su \cite{MS08}.  In the MSSM a radiative correction to the beta-decay amplitude involves
box graphs with exchanged supersymmetric sfermions.  These produce an energy dependence in the
beta-decay amplitude that shows up in the Fierz interference term.  Unfortunately, estimates by Profumo, 
Ramsey-Musolf and Tulin \cite{PRT07} indicate that the resulting value of $b_F$ would be less than
$10^{-3}$, which is an order of magnitude smaller than our current experimental upper limit on that
quantity.  Thus it will likely be some time before superallowed beta decay can provide useful
constaints for this class of supersymmetric models.  

\subsubsection{ Induced Scalar current}
\label{sss:isc}

If we consider only the vector part of the weak interaction, for composite
spin-1/2 nucleons the
most general form of that interaction is written \cite{BB82} as 
\be
H_V = \overline{\psi}_p ( \gV \gamma_{\mu} - \fM \sigma_{\mu \nu} q_{\nu}
+ i \fS q_{\mu} ) \psi_n ~ \overline{\phi}_e \gamma_{\mu} ( 1 +
\gamma_5 ) \phi_{\overline{\nu}_e},
\label{HVnucl}
\ee
with $q_{\mu}$ being the four-momentum transfer, $q_{\mu} = (p_p - p_n)_{\mu}$.
The values of the coupling constants $\gV$ (vector), $\fM$ (weak magnetic) and
$\fS$ (induced scalar) are prescribed so long as the CVC hypothesis -- that the
weak vector current is just an isospin rotation of the electromagnetic vector
current -- is correct.  In particular, since CVC implies that the vector current
is divergenceless, it follows that $\fS$ should equal zero.  An independent
argument \cite{We58}, that there be no second-class currents in the hadronic
weak interaction, also requires $\fS$ to vanish.  We proved in our previous
survey \cite{HT05} that the presence of a non-zero $\fS$ would manifest itself
in exactly the same way as a non-zero $C_S$: by a $\langle1/W\rangle$
dependence in the $\F t$-value data.

In the same manner that we obtained Eqs.~(\ref{bF}) and (\ref{CS/CV}), we determine
from our present survey results that
\be
m_e \fS / \gV = - ( 0.0011 \pm 0.0013) .
\label{fsgv}
\ee
This result is a vindication for the CVC hypothesis,
which predicts $\gV = 1$ and $\fS = 0$.  We confirm this prediction at the level of 24
parts in $10^4$.  Our result can also be interpreted as setting a limit on vector
second-class currents in the semi-leptonic weak interaction.

\subsection{Limits on extensions to the standard model}
\label{ss:limits}

The unitarity sum established in Sect.~\ref{ss:unitarity} can be used to set limits on new
physics beyond the standard model.  A list of possible extensions includes, but is certainly
not limited to, right-hand currents, extra $Z$ bosons, scalars, supersymmetry, a fourth
generation of quarks and exotic muon decay.  Marciano surveyed many of these possibilities
at the CKM2008 Workshop \cite{CKM08}.  His general conclusion was that although the CKM
unitarity test yields no sign of new physics, it does place important constraints on the
possibilities.  In the case of supersymmetric models these constraints have been explored
by Ramsey-Musolf, Su and Kurylov \cite{RS08,RM00,KRM02}.  
In the minimal supersymmetric version (MSSM), corrections
to low-energy observables arise only via loop effects, while in extensions that allow for
$R$-parity violating (RPV) interactions new tree-level effects appear. In general, the
presence of new physics may modify low-energy semi-leptonic electroweak observables in two
ways: {\it (i)} directly, via a new semi-leptonic interaction (\eg right-hand currents, MSSM
with RPV interactions), and {\it (ii)} indirectly, via loop graphs contributing to the
radiative correction (\eg extra $Z$-bosons, MSSM).  In what follows we give one example of
each type of modification: right-hand currents and extra $Z$ bosons.

\subsubsection{Right-hand currents}
\label{sss:rhc}

In the standard model, parity violation is considered to be maximal.  What if this
condition were to be relaxed?  For semi-leptonic transitions, Herczeg \cite{He86,He01a}
extends the general form of the weak interaction to read
\be
H_{s \ell} = a_{LL}(V-A)(V-A) + a_{LR}(V-A)(V+A) + a_{RL}(V+A)(V-A)
             + a_{RR}(V+A)(V+A) ,
\label{Hsl}
\ee
where, in each term, the first factor represents the lepton currents and the second, 
the hadron currents.  In particular, for the vector lepton current, $V$ stands for either
$\overline{\phi}_e \gamma_{\mu} \phi_{{\overline{\nu}}_e}^L$ or
$\overline{\phi}_e \gamma_{\mu} \phi_{{\overline{\nu}}_e}^R$ depending on whether the
chirality of the neutrino is left-handed, as it is for $V-A$ coupling, or right-handed, 
as it is for $V+A$ coupling.  In the standard model, $a_{LL} = 1$, and
$a_{LR} = a_{RL} = a_{RR} = 0$.  For Fermi beta decay, only the vector part of the weak
hadron current contributes, so the decay rate is given by the following proportionality
\cite{HT05}:
\bea
\Gamma_{\beta} & \propto & |a_{LL} + a_{LR} |^2 + |a_{RL} + a_{RR} |^2
\nonumber \\
& \simeq & |a_{LL}|^2 \left ( 1 + 2 Re \overline{a}_{LR} + \ldots \right ) ,
\label{Gammb}
\eea
where $\overline{a}_{LR} = a_{LR}/a_{LL}$.  In the second line of the equation, we have
only retained quantities that are first order in the (presumably) small quantities
$\overline{a}_{LR}$, $\overline{a}_{RL}$ and $\overline{a}_{RR}$.

To determine the effect that right-hand currents would have on the value of $V_{ud}$ obtained
from experiment, we also need to consider the role of such currents on the purely leptonic muon
decay.  Herczeg \cite{He86} writes the effective Hamiltonian, in analogy to Eq.~(\ref{Hsl}), as
\be
H_{\ell} = c_{LL}(V-A)(V-A) + c_{LR}(V-A)(V+A) + c_{RL}(V+A)(V-A)
+ c_{RR}(V+A)(V+A).
\label{Hl}
\ee
The coupling constants in Eqs.~(\ref{Hl}) and (\ref{Hsl}) are related by the CKM matrix elements:
\bea
a_{LL} = c_{LL} V_{ud}^L & & a_{LR} = c_{LR} e^{i \alpha} V_{ud}^R
\nonumber \\
a_{RL} = c_{RL} V_{ud}^L & & a_{RR} = c_{RR} e^{i \alpha} V_{ud}^R .
\label{ac}
\eea
Here $V_{ud}^L$ is the element of the CKM matrix for left-handed chirality quarks, and $V_{ud}^R$
is for right-handed chirality quarks.  The phase $\alpha$ is a $CP$-violating phase in the
right-handed CKM matrix.  The decay rate for muon decay is constructed from an equal mix of
vector and axial-vector interactions and is proportional to the following expression \cite{HT05}:
\bea
\Gamma_{\mu} & \propto & |c_{LL}|^2 + |c_{LR}|^2 + |c_{RL}|^2 + |c_{RR}|^2
\nonumber \\
& = & |c_{LL}|^2 \left ( 1 + |\overline{c}_{LR}|^2 + |\overline{c}_{RL}|^2
+ | \overline{c}_{RR}|^2 \right ) ,
\label{Gammu}
\eea
where $\overline{c}_{ij} = c_{ij}/c_{LL}$.

If we define $|V_{ud}|^2_{expt}$ as being the quantity obtained from the ratio of
measured beta- and muon-decay rates, we can combine Eqs.~(\ref{Gammb}) and (\ref{Gammu})
to relate this experimental result to the matrix element $|V_{ud}^L|^2$ by the relationship  
\bea
|V_{ud}|^2_{expt} \equiv  \frac{\Gamma_{\beta}}{\Gamma_{\mu}} & = &
|V_{ud}^L|^2 \frac{|1 + \overline{a}_{LR}|^2 + |\overline{a}_{RL} +
\overline{a}_{RR}|^2}{1 + |\overline{c}_{LR}|^2 + |\overline{c}_{RL}|^2
+ |\overline{c}_{RR}|^2}
\nonumber \\
& \simeq & |V_{ud}^L|^2 ( 1 + 2 Re \overline{a}_{LR} ) ,
\label{Gammbu}
\eea
where, in the second line, only corrections to first order in small quantities are
retained.  If the situation is identical for the second (kaon decay) and third
(B-meson decay) generations of quarks, with the interaction coupling constants $a_{ij}$
and $c_{ij}$ in $H_{s \ell}$ and $H_{\ell}$ being generation independent, then
\bea
\sum_i |V_{ui}|_{\rm expt}^2 & = & \sum_i |V_{ui}^L|^2
( 1 + 2 Re \overline{a}_{LR} )
\nonumber \\
& = & 
 1 + 2 Re \overline{a}_{LR} .
\label{unitaLR}
\eea
In writing the second line we have assumed that the CKM matrix for left-hand chirality
quarks is strictly unitary.  Since the left-hand side of Eq.~(\ref{unitaLR}) is the
experimentality determined unitarity sum, given in Eq.~(\ref{usum}) of Sect.~\ref{ss:unitarity}, 
this expression can clearly be used to set a limit on the coupling constant $\overline{a}_{LR}$.
The result is
\bea
0.99995 \pm 0.00061 & = &
 1 + 2 Re \overline{a}_{LR} 
\nonumber \\
 Re \overline{a}_{LR} & = & -0.00003 \pm 0.00030 ,
\label{norhc}
\eea
which is consistent with no right-hand currents -- at least not in the LR sector.

\subsubsection{Extra $Z$ bosons}
\label{sss:extraZ}

The existence of neutral gauge bosons, beyond the usual photon and $Z$ boson of the standard
$SU(2)_L \times U(1)$ model, would impact on the CKM unitarity test.  To illustrate this we
consider just one of the many models that appear in grand unified theories, namely the
$SO(10)$ model, whose group breakdown is
\be
SO(10) \rightarrow SU(3)_C \times SU(2)_L \times U(1) \times U(1)_{\chi} .
\label{SO10}
\ee
Here an extra $U(1)$ group is introduced, $U(1)_{\chi}$, and its corresponding neutral gauge
boson is labelled $Z_{\chi}$.  The existence of such an extra $Z$ boson would impact on the
calculation of the electroweak radiative correction.  One of the important diagrams in the
hadron-independent radiative correction, $\DRV$, is a $WZ$-box graph.  This graph would have
to be augmented by an additional $WZ_{\chi}$-box graph, whose contribution is of order
$\ln x_{\chi}$, where $x_{\chi} = m_{Z_{\chi}}^2/m_W^2$, the ratio of squared masses of the
heavy bosons in the box diagram.  If we assume that this correction is common to all quark
flavours, then the same correction that occurs in the determination of $|V_{ud}|^2$ would
also occur for $|V_{us}|^2$ and $|V_{ub}|^2$.  If so, its impact can be incorporated into the 
unitarity test.

Following Marciano and Sirlin \cite{MS87}, we write
\be
|V_{ud}|^2 + |V_{us}|^2 + |V_{ub}|^2 =
(0.99995 \pm 0.00061) + \Delta ,
\label{Delta}
\ee
where the numerical value for the experimental unitarity sum is from Eq.~(\ref{usum}), and
$\Delta$ is a calculated correction due to the extra $Z$ boson.  If we take the CKM matrix
to be exactly unitary in three generations, then Eq.~(\ref{Delta}) can be used to set the
following one-standard-deviation limits on $\Delta$:
\be
-0.00056 \leq \Delta \leq +0.00066 .
\label{Dlimit}
\ee
Marciano and Sirlin \cite{MS87} have computed the contribution of a putative $Z_{\chi}$ boson
to the radiative correction and obtained
\be
\Delta = - \frac{27 \alpha}{40 \pi \sin^2 \theta_W }
\times \frac{4}{3} |C_{\chi}|^2 \frac{\ln x_{\chi}}{x_{\chi}-1} ,
\label{Dvalu}
\ee
where $\alpha$ is the fine-structure constant, $\theta_W$ is the Weinberg angle
($\sin^2 \theta_W \simeq 0.23$), and $C_{\chi}$ is a coupling constant linking the $Z_{\chi}$
boson to fermions.  The normalization has been selected so that $C_{\chi}$ is unity at the
$SO(10)$ unification mass scale.  Its value at lower energies has to be estimated, and
Marciano and Sirlin use $|C_{\chi}|^2 = \sfrac{1}{2}$.  Noting that the correction $\Delta$
is negative, we obtain from the lower limit in Eq.~(\ref{Dlimit}) 
\be
\frac{\ln x_{\chi}}{x_{\chi}-1} \leq 0.12 .
\label{xlimit}
\ee
Taking for the $W$-boson mass, $m_W = 81$ GeV, we arrive at the limit
\be
m_{Z_{\chi}} > 430~{\rm GeV} .
\label{mlimit}
\ee

Impressive though this limit is, somewhat higher limits have been obtained in direct searches
at proton and electron colliders.  The CDF and D0 experiments at FermiLab in searches of
$\overline{p}p \rightarrow e^+ e^-$ have placed lower-mass limits (at 95\% C.L.) on $m_{Z_{\chi}}$
of 822 and 640 GeV respectively, while at CERN the LEP2 experiment on $e^+ e^- \rightarrow
f \overline{f}$ (with $f$ signifying a fermion) find a lower-mass limit of 673 GeV.  
These limits are recorded in the survey of Erler and Langacker \cite{EL08} in the 2008 Particle
Data Group listings.

\section{\label{s:concl} Conclusions}

In our previous survey \cite{HT05}, only four years ago, we remarked on the excellent agreement
among the derived $\F t$ values, lamented that the results of the unitarity test were still ambiguous, 
and predicted that the already well-measured $ft$ values of the ``traditional nine" superallowed
decays were unlikely to be improved dramatically in the near future.  Much has happened since then, not
all of it expected.  Today, we can say that the excellent $\F t$-value consistency remains -- or, to
be more accurate, it has been restored after Penning-trap $Q_{EC}$-value measurements, non-existant
at the time of the last survey, did in fact make important improvements (and changes) in the known
$ft$ values, which in turn prompted improvements (and changes) in the calculated isospin-symmetry-breaking
corrections.  At the same time, the calculation of the nucleus-independent radiative correction, $\DRV$, was
improved, leading to a more precise result for $V_{ud}$, and the kaon-decay community mounted a concerted
effort, which led to a new and reliable value for $V_{us}$.  With these new results, and others,
CKM unitarity has now been tested to unprecedented precision \ldots and it has passed the test with flying
colors.

Furthermore, we have demonstrated in Sec.~\ref{s:impact} how powerful these improved results can be in setting
limits on new physics beyond the standard model, whether that new physics be a scalar interaction, right-hand
currents or extra $Z$ bosons.  We have seen that tiny uncertainties on the $ft$ values are essential
ingredients of a demanding test of CKM unitarity, which also leads to tight limits on new physics.  The
challenge now is: Can those uncertainties be reduced still farther?  The motivation is as strong as ever: to
identify the need for new physics -- or to limit the possible candidate theories even more definitively.

We have taken pains throughout this work to pay careful attention to all uncertainties, theoretical
and experimental.  In Sec.~\ref{ss:vVud} we detail the various contributions to the uncertainty in
$|V_{ud}|^2$.  Of these, by far the largest is still from $\DRV$, even though its uncertainty has
recently been improved significantly \cite{Ma06}.  To improve it more must remain an important theoretical
goal.

The next largest contributor to the error budget for $|V_{ud}|^2$ is the nuclear-structure-dependent
corrections, ($\delta_C - \delta_{NS}$).  Their uncertainties arise both from the input parameters used
in their calculation -- two-body matrix elements in the shell-model calculations, experimental
uncertainties in charge-radii, etc \cite{TH02,TH08} -- and from possible systematic differences between
two different methods used for calculating radial wave functions (see Sec.~\ref{ss:se}).  From a theoretical
point of view, it would obviously be desirable to have a third completely different calculation, to
reinforce the assessment of systematic uncertainties.  However, in the absence of such a calculation,
one must rely on experiment to test the accuracy of these calculated corrections. This has become, and
should remain, a top priority for experiment.

\begin{figure}[t]
\epsfig{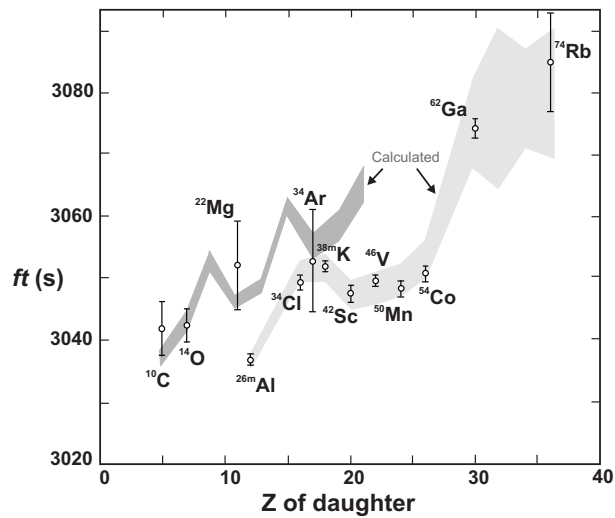}
\caption{Experimental $ft$ values plotted as a function of the charge on the daughter nucleus, $Z$.  
Both bands represent the quantity $\overline{\F t}/((1 + \delta_R^{\prime})(1 - \delta_C + \delta_{NS}))$. 
The two separate bands distinguish those beta emitters whose parent nuclei have isospin $T_z = -1$
(darker shading) from those with $T_z =0$ (lighter shading).}
\label{fe:dcvar2}
\end{figure}

The method, which is best described with reference to Fig.~\ref{fe:dcvar2}, is based on the validity
of the CVC hypothesis that the corrected $\F t$ values for the superallowed $0^+ \rightarrow
0^+$ decays should be constant.  In the figure we compare the uncorrected measured $ft$ values
(points and error bars) with the quantity $\overline{\F t}/((1 + \delta_R^{\prime})(1 - \delta_C
+ \delta_{NS}))$ shown as a band, the width of which represents the assigned theory error.  The band
corresponds to the calculated corrections normalized to the data via the measured average
$\F t$ value, $\overline{\F t}$, taken from Table~\ref{Ft}.  Thus, although this comparison does
not test the absolute values of the correction terms, it does test the collective ability of all
three calculated correction terms to reproduce the significant variations in $ft$ from one
transition to another.  In fact, since $\delta_R^{\prime}$ is almost independent
of $Z$ when $Z>10$, this test really probes directly the effectiveness of the calculated values
of ($\delta_C-\delta_{NS})$.   

It can be seen that there is remarkable agreement between theory and experiment.  In assessing
the significance of this agreement, it is important to recognize that the calculations of
$\delta_C$ and $\delta_{NS}$ for $Z\leq 26$ are based on well-established shell-model wave
functions that were further tuned to reproduce measured binding energies, charge radii and
coefficients of the isobaric multiplet mass equation \cite{TH02,TH08}.  The origins of the
calculated correction terms for all cases are completely independent of the superallowed
decay data.  Thus, the agreement in the figure between the measured superallowed data points
and the theoretical band is already a powerful validation of the calculated corrections used
in determining that band.  The validation becomes even more convincing when we consider that it
would require a pathological fault indeed in the theory to allow the observed nucleus-to-nucleus
variations in $\delta_C$ and $\delta_{NS}$ to be reproduced in such detail while failing to
obtain the {\em absolute} values to comparable precision.  As satisfactory as the agreement in
Fig.~\ref{fe:dcvar2} is, though, new experiments can still improve the test, making it even more
demanding, and can ultimately serve to reduce the uncertainty in the nuclear-structure-dependent
corrections even further.

These new experiments can follow different paths.  In the last four years, the biggest impact has
come from experiments that focused on the ``traditional nine" superallowed transitions.
New Penning-trap $Q_{EC}$-value measurements have already been mentioned, but there have been new
half-life and branching-ratio measurements as well (see Tables~\ref{t1/2} and \ref{R}).  More
improvements are still possible as a glance at Fig.~\ref{f:hist9} reveals.  If we accept as a goal
that experiment should be more than a factor of two more precise than theory, then we see that the
$Q_{EC}$ values for $^{10}$C, $^{14}$O and $^{34}$Cl, the half-lives of $^{26}$Al$^m$, $^{34}$Cl,
$^{42}$Sc and $^{50}$Mn, and the branching ratios for $^{10}$C and $^{14}$O can all bear improvement.
It is also particularly noteworthy that any improvements in the cases of $^{10}$C and $^{14}$O will
lead directly to improvements in the limits on the possible existence of scalar currents.  As is
evident from Fig.~\ref{f:scalar} and the discussion in Sec.~\ref{sss:fsc}, it is on these two low-$Z$
superallowed transitions that a scalar current would have the largest effect.  Unfortunately the
branching ratios for both these transitions offer experimental obstacles that have proved very
difficult to surmount.

A second experimental path is to expand the number of precisely measured superallowed emitters to
include cases for which the calculated nuclear-structure-dependent corrections are larger, or show
larger variations from nuclide to nuclide, than the values applied to the ``traditional nine" cases.
We argue that if the experimental $ft$ values agree with the calculations where the nucleus-to-nucleus
variations are large, then that must surely verify the calculations' reliability for the nine cases
whose corrections are considerably smaller.  Already four cases of this type have been carefully
measured, $^{22}$Mg, $^{34}$Ar, $^{62}$Ga and $^{74}$Rb.  They appear to agree well with the
calculations although, with the exception of $^{62}$Ga, their uncertainties are still five times
greater than those for the best known transitions.  Undoubtedly these uncertainties will be reduced
and more cases added in the near future.

These new cases certainly present serious experimental challenges.  The parent nuclei are more exotic
than the traditional cases, which all have stable daughters, so they are more difficult to produce in
pure and statistically significant quantities.  They also exhibit more complex branching patterns: Each 
$T_Z\,$=$\,-1$ parent nucleus decays by Gamow-Teller transitions of comparable strength to the superallowed
Fermi one, thus requiring the latter's branching ratio to be measured directly with high precision.
For the $T_Z\,$=$\,0$ parents with $A\ge62$, each decay includes numerous weak Gamow-Teller transitions, 
which are very difficult to observe individually but which collectively constitute nonnegligible
branching strength.  In both regions, these problems are being, or have been overcome, albeit with very
specialized techniques.  The recently published branching-ratio measurement \cite{Fi08} for $^{62}$Ga is
an example of how even meticulously detailed spectroscopic studies must be combined with theory
\cite{Ha02} to ensure that missing transitions are properly accounted for in the decays of the heavy
$T_Z\,$=$\,0$ parents.

There is a further important issue that arises for the superallowed emitters with $A\ge62$: The
shell-model calculations of the structure-dependent corrections for these nuclei are not solidly based
on spectroscopic measurements as they are for the lighter nuclei.  Such measurements simply do not
exist for most $N\simeq Z$ nuclei in this mass region.  Furthermore, charge radii and coefficients
for the isobaric multiplet mass equation are not known either and so cannot be used to constrain the
radial wave functions or ``tune" the charge-dependence embedded in the two-body matrix elements.  As
a consequence, the uncertainties assigned to the calculated corrections are very large (see the
broad band in this mass region in Fig.~\ref{fe:dcvar2}), considerably reducing the usefulness of
these nuclei either in testing the corrections or in contributing to the determination of $V_{ud}$.  
It would be very valuable in this context for radioactive-beam facilities to direct some attention
to determining a wide variety of spectroscopic information in this mass region with a view to
obtaining a reasonably effective nuclear model, which, among other things, could lead to much improved
calculations for the correction terms.

In conclusion, we can assert -- as we did four years ago -- that world data for superallowed
$0^+$$\rightarrow 0^+$ $\beta$ decays strongly support the CVC expectation of an unrenormalized vector
coupling constant, and also set a tight limit, consistent with zero, on scalar currents.  We
can now add, though, that CKM unitarity is satisfied to within an uncertainty of 0.06\%.  This
reconciliation with unitarity has come about as a result of significant changes in $V_{us}$; the value
of $V_{ud}$ determined from nuclear $\beta$ decay has not varied outside of error bars in twenty
years, during which time the size of those error bars has been reduced by a factor of five.  Finally,
we have noted that the calculated nuclear-structure dependent correction terms have recently been 
improved and continue to stand up favorably to experimental tests, an outcome that must further
increase confidence in the nuclear results.

\acknowledgments

We wish to thank Wick Haxton for prompting us to include atomic effects in the calculation
of $f$ and for suggesting how we could go about doing it.  We also appreciate helpful
correspondence from Erich Ormand and Michael Ramsey-Musolf.  The work of J.C.H. was supported by
the U. S. Dept.~of Energy under Grant DE-FG03-93ER40773 and by the Robert A. Welch Foundation
under Grant A-1397.  I.S.T. would like to thank the Cyclotron Institute of Texas A \& M University
for its hospitality during annual two-month visits.

\appendix
\section{Atomic Overlap Correction to the Statistical Rate Function}
\label{s:srf}

The statistical rate function, $f$, is an integral over phase space,
\be
f = \int_1^{W_0} p W (W_0 - W)^2 F(Z,W) S(Z,W)~dW ,
\label{fold}
\ee
where $W$ is the total energy of the electron in electron-rest-mass units;
$W_0$ is the maximum value of $W$; $p = (W^2 - 1)^{1/2}$ is the momentum
of the electron; $Z$ is the atomic number of the daughter nucleus; $F(Z,W)$ is
the Fermi function and $S(Z,W)$ is the shape-correction function.  The details
of the calculation of $S(Z,W)$ were given in our previous survey \cite{HT05}
and will not be repeated here.  What we address here is the inclusion for the first
time of an additional factor in Eq.~(\ref{fold}) to account for the mismatch in the
initial and final {\it atomic} states in the $\beta$ decay.  Since the nucleus
changes charge by one unit in beta decay, the final atomic state does not overlap
perfectly with the initial atomic state, an effect that leads to a slight
inhibition in the beta-decay rate.  In the past, this effect has justifiably been
considered too small to be of practical concern but, with the advent of Penning-trap
mass measurements, the experimental uncertainties in transition $Q$-values have
been reduced so much that they are now comparable to the effects of the imperfect
atomic overlap. 

We begin by writing
\be
f = \int_1^{W_0} p W (W_0 - W)^2 F(Z,W) S(Z,W) r(Z,W)~dW ,
\label{fnew}
\ee
where $r(Z,W)$ is the atomic overlap correction we are seeking.
We then follow the method of Bahcall \cite{Ba63} by expressing $f$ as a
double integral with an energy-conserving delta function:
\be
f = \int \int pWq^2 F(Z,W) S(Z,W) \sum_{A^{\prime}} | \langle A^{\prime} |
G \rangle |^2 \delta(E_f - E_i) ~ dW dq ,
\label{fdouble}
\ee
where $q$ is the neutrino momentum.  We have introduced into this equation an
overlap of the initial and final atomic electron configurations:
$|G\rangle$ is the state vector for the initial neutral
atom with $(Z+1)$ electrons, and $|A^{\prime}\rangle$ is the
state vector for the final {\em ionized} atom with $(Z+1)$ electrons
but only charge $Z$ in the nucleus.  There are many such possible
final states, so a sum over $A^{\prime}$ is included.

Of the two energies within the delta function, the first, $E_i$, is the energy
of the initial neutral atom in its atomic ground state:
\be
E_i = {\cal M}_{Z+1}(G) = M_{Z+1} + (Z+1)m_e - B(G) ,
\label{Ei}
\ee
where ${\cal M}_{Z+1}(G)$ is the {\em atomic} mass, $M_{Z+1}$ is
the {\em nuclear} mass, $m_e$ is the electron mass ($m_e = 1$ in
electron rest-mass units) and $B(G)$ is the total electron
binding energy in the ground state of the atom.  The sign of the latter is 
chosen so that $B(G) > 0$.  The second energy, $E_f$, is that of the
final state, which is composed of an {\em ionized} atom still with $(Z+1)$
atomic electrons in an excited configuration plus an
emitted beta-decay positron and an emitted neutrino:
\be
E_f = {\cal M}_{Z+1}^{-1}(A^{\prime}) + W + q =
M_Z + (Z+1)m_e - B(A^{\prime}) + W + q ,
\label{Ef}
\ee
where ${\cal M}_{Z+1}^{-1}(A^{\prime})$ is the atomic mass of a
negatively ionized atom (superscript $-1$
denotes ionization) of $(Z+1)$ electrons
in configuration $A^{\prime}$, $M_Z$ is the nuclear mass for the final nucleus
in the beta decay, and $B(A^{\prime})$ is the total electron 
binding energy for the ionized atom.  Thus the energy difference
becomes
\bea
E_f - E_i & = & (M_Z - M_{Z+1}) + W + q +B(G) - B(A^{\prime})
\nonumber \\
& = & (M_Z - M_{Z+1}) + W + q +[B(G) - B(G^{\prime})]
-[B(A^{\prime}) - B(G^{\prime})] .
\label{EfEi}
\eea
In the second line of the equation we have introduced the total electron binding
energy for the final {\em neutral} atom of charge $Z$ in its atomic 
ground-state configuration, $B(G^{\prime})$.

The $Q_{EC}$ value is the difference in the atomic masses
of neutral atoms in ground-state configurations:
\bea
Q_{EC} & = & {\cal M}_{Z+1}(G) - {\cal M}_Z(G^{\prime})
\nonumber \\
& = & [M_{Z+1} + (Z+1)m_e - B(G)] - [M_Z + Z m_e - B(G^{\prime})]
\nonumber \\
& = & (M_{Z+1} - M_Z ) + m_e + [B(G^{\prime}) - B(G)] ;
\label{Qec}
\eea
and the quantity $W_0$ in Eq.~(\ref{fold}) is related to $Q_{EC}$ by the equation
$W_0 = Q_{EC} - m_e$.  Thus, $E_f - E_i$ in Eq.~(\ref{EfEi})
can be written as 
\be
E_f - E_i = q + W - W_0 + [B(G^{\prime}) - B(A^{\prime})] .
\label{EfEi1}
\ee
For the energy-conserving delta function we now make a Taylor
series expansion about the value $q + W - W_0$:
\be
\delta (E_f - E_i) = \delta ( q+W-W_0) + \delta^{\prime}(q+W-W_0)
[B(G^{\prime}) - B(A^{\prime})] + \ldots
\label{dEfEi}
\ee
If the first term in this expansion is inserted into the double integral,
Eq.~(\ref{fdouble}), then the expression for $f$ reduces to the
original form Eq.~(\ref{fold}) since the atomic overlap factor is
unity under the assumption that the sum over electronic 
configurations $A^{\prime}$ can be completed by closure: {\it i.e.}
$\sum_{A^{\prime}} | \langle A^{\prime} | G \rangle |^2 =
\sum_{A^{\prime}} \langle G | A^{\prime} \rangle \langle A^{\prime} | G
\rangle = \langle G | G \rangle = 1$.  The second term in Eq.~(\ref{dEfEi})
involves a derivative of a delta function.  This is handled by
an integration by parts, in which the rest of the integrand is
differentiated with respect to $q$.  No boundary terms
survive as the integrand vanishes at the boundaries.
Thus the atomic overlap correction becomes
\bea
r(Z,W) & = & 1 - \frac{2}{W_0-W} \sum_{A^{\prime}}
| \langle A^{\prime} | G \rangle |^2 \left [
B(G^{\prime}) - B(A^{\prime}) \right ]
\nonumber \\ 
& = & 1 - \frac{2}{W_0-W} \left ( B(G^{\prime}) - \sum_{A^{\prime}}
| \langle A^{\prime} | G \rangle |^2 
B(A^{\prime}) \right ) .
\label{r1}
\eea

Next, it is useful to recall the eigenvalue equations satisfied by
the atomic states $|G\rangle$ and $|A^{\prime}\rangle$:
\be
\hat{H}_i |G\rangle = - B(G) |G\rangle ~~ {\rm with} ~~
\hat{H}_i = \sum_{i=1}^{Z+1} t_i - (Z+1)e^2 \sum_{i=1}^{Z+1} \frac{1}{r_i}
+ e^2 \sum_{i<j=1}^{Z+1} \frac{1}{r_{ij}} ,
\label{HG}
\ee
\be
\hat{H}_f^{-1} |A^{\prime}\rangle = - B(A^{\prime}) |A^{\prime}\rangle ~~ {\rm with} ~~
\hat{H}_f^{-1} = \sum_{i=1}^{Z+1} t_i - Ze^2 \sum_{i=1}^{Z+1} \frac{1}{r_i}
+ e^2 \sum_{i<j=1}^{Z+1} \frac{1}{r_{ij}} ,
\label{HAprime}
\ee
where $t_i$ is the kinetic energy of electron $i$, $r_i$ is its distance
from the nucleus and $r_{ij}$ is the separation of electrons $i$ and $j$.
Note, in particular, that
\be
\hat{H}_i - \hat{H}_f^{-1} = -e^2 \sum_{i=1}^{Z+1} \frac{1}{r_i}
\simeq
\frac{\partial}{\partial Z} \hat{H}_i .
\label{Hdiff}
\ee
Inserting these Hamiltonian expressions into Eq.~(\ref{r1}) we obtain
\bea
r(Z,W) & = & 1 - \frac{2}{W_0-W} \left ( B(G^{\prime}) +
\sum_{A^{\prime}} \langle G | \hat{H}_f^{-1} | A^{\prime} \rangle
\langle A^{\prime} | G \rangle \right )
\nonumber \\
& = & 1 - \frac{2}{W_0-W} \left ( B(G^{\prime}) +
\langle G | \hat{H}_f^{-1} | G \rangle \right )
\nonumber \\
& = & 1 - \frac{2}{W_0-W} \left ( B(G^{\prime}) -
\langle G | \hat{H}_i - \hat{H}_f^{-1} | G \rangle 
+ \langle G | \hat{H}_i | G \rangle \right )
\nonumber \\
& = & 1 - \frac{2}{W_0-W} \left ( B(G^{\prime}) +
\frac{\partial}{\partial Z} B(G) - B(G) \right ) ,
\label{r2}
\eea

\begin{table}[t]
\begin{center}
\caption{Comparison of statistical rate functions calculated without the atomic overlap
correction, $f_{{\rm without}}$, those calculated with it included, $f_{{\rm with}}$.
The change in the $Q_{EC}$ value that would lead to the same change
in $f$ is given in the last column. 
\label{t:overlap}}
\vskip 1mm
\begin{ruledtabular}
\begin{tabular}{rrrrrr}
& & & & & \\[-3mm]
Parent & $f_{\rm without}$ & $f_{\rm with}$ & $df/f(\%)$ & $dQ/Q(\%)$ & 
$dQ$ (eV)
\\[1mm]
\hline
& & & & & \\[-3mm]
$T_z = -1$: & & & & & \\
$^{10}$C  &  2.30089 &  2.30039 & 0.02178 & 0.00436 & 83 \\
$^{14}$O  &  42.7779 &  42.7724 & 0.01277 & 0.00255 & 72 \\
$^{18}$Ne &  134.484 &  134.469 & 0.01093 & 0.00219 & 74 \\
$^{22}$Mg &  418.423 &  418.386 & 0.00877 & 0.00175 & 72 \\
$^{26}$Si &  1029.52 &  1029.44 & 0.00767 & 0.00153 & 74 \\
$^{30}$S  &  1967.05 &  1966.91 & 0.00707 & 0.00141 & 77 \\
$^{34}$Ar &  3414.68 &  3414.46 & 0.00647 & 0.00129 & 78 \\
$^{38}$Ca &  5327.57 &  5327.24 & 0.00612 & 0.00122 & 81 \\
$^{42}$Ti &  7040.63 &  7040.21 & 0.00597 & 0.00119 & 84 \\[5mm]
$T_z = 0 $: & & & & & \\
$^{26m}$Al&  478.279 &  478.237 & 0.00880 & 0.00176 & 75 \\
$^{34}$Cl &  1996.10 &  1995.96 & 0.00711 & 0.00142 & 78 \\
$^{38m}$K &  3298.10 &  3297.88 & 0.00663 & 0.00133 & 80 \\
$^{42}$Sc &  4472.52 &  4472.24 & 0.00643 & 0.00129 & 83 \\
$^{46}$V  &  7209.90 &  7209.47 & 0.00598 & 0.00120 & 84 \\
$^{50}$Mn &  10746.6 &  10746.0 & 0.00565 & 0.00113 & 86 \\
$^{54}$Co &  15767.5 &  15766.6 & 0.00537 & 0.00107 & 89 \\
$^{62}$Ga &  26401.6 &  26400.2 & 0.00557 & 0.00111 &102 \\
$^{66}$As &  32127.0 &  32125.3 & 0.00545 & 0.00109 &104 \\
$^{70}$Br &  38602.2 &  38600.1 & 0.00539 & 0.00108 & 107 \\
$^{74}$Rb &  47296.9 &  47294.5 & 0.00423 & 0.00105 & 109 \\
\end{tabular}
\end{ruledtabular}
\end{center}
\end{table}

\noindent
where in the last line the order of integration and differentiation has been
reversed on the assumption that the binding energy as a function of $Z$
behaves in a smooth way.  Now, $B(G^{\prime})$ is the electronic
binding energy of a neutral atom with $Z$ electrons, while $B(G)$ is the 
same quantity for an atom with $Z+1$ electrons.  Treating $B(G^{\prime})$
as a function of $Z$, we can expand $B(G^{\prime})$ in a Taylor series
about $B(G)$:
\be
B(G^{\prime}) = B(G) - \frac{\partial}{\partial Z}B(G) +
\frac{1}{2} \frac{\partial^2}{\partial Z^2} B(G) - \ldots
\label{taylor}
\ee
Then, substituting this expression in Eq.~(\ref{r2}), we obtain our final expression
for the atomic overlap correction:
\be 
r(Z,W) = 1 - \frac{1}{W_0-W} \frac{\partial^2}{\partial Z^2} B(G) .
\label{r3}
\ee
This expression was first obtained by Bahcall \cite{Ba63}.

It remains for us to estimate the second derivative of the electronic binding
energy of neutral atoms in their ground-state configuration.  For this we use
binding-energy values from the tables of Carlson \etal \cite{Ca70}, which were obtained
from self-consistent Hartree-Fock calculations and have been demonstrated to agree
with experimental values to within 5\%.  We performed a fit to these tabulated values
using a fitting function, $a Z^b$, in three ranges of $Z$ values, with the following
results:
\be
B(G) = \begin{array}{ll}
13.080 Z_i^{2.42}~{\rm eV}, & ~~~~6 \leq Z_i \leq 10 \\[2mm]
14.945 Z_i^{2.37}~{\rm eV}, & ~~~11 \leq Z_i \leq 30 \\[2mm]
11.435 Z_i^{2.45}~{\rm eV}, & ~~~31 \leq Z_i \leq 39 ,
\end{array}
\label{BG}
\ee
where $Z_i$ is the charge of the parent atom in the beta-decay process.
It is conventional to use $Z$ as the charge of the daughter nucleus
in beta decay; thus for positron decay $Z_i = Z+1$.  The second derivative 
is easily obtained from these expressions.

We have re-computed the statistical rate function $f$, with the results being
listed in Table~\ref{t:overlap}.  Those results obtained without
the atomic overlap correction, Eq.~(\ref{fold}), are given under
the heading $f_{\rm without}$, while those with the correction, 
Eqs.~(\ref{fnew}) and (\ref{r3}), are labelled $f_{\rm with}$.  The latter
results also appear in column 2 of Table~\ref{Ft}.  The fractional difference
between $f_{\rm with}$ and $f_{\rm without}$ in percent is given in
column 4 and is of order 0.01\%, decreasing with increasing
mass value.  This is a very small correction.  Furthermore, the statistical rate
function depends on the $Q$-value to the fifth power, so the fractional change in $Q$
that would lead to a change in $f$ of the same size as that induced by the atomic
overlap correction is even smaller: $1/5 \times df/f$.  This percentage change is
given in column 5 of Table~\ref{t:overlap}. As small as this effect is, it can be
seen from the last column of the table that the equivalent change in $Q$-value ranges
from 70 to 110 eV, an amount that is similar to the experimental uncertainties on
the most precisely measured $Q$-values.

\section{Accounting for systematic errors}
\label{s:syserror}
\subsection{Isospin-symmetry-breaking correction, $\delta_{C2}$}
\label{ss:deltac}
 
In the past (see, for example, \cite{HT05}), we have added a systematic uncertainty
to the average corrected $\overline{\F t}$ value to account for an apparent systematic
difference between calculations of the isospin-symmetry-breaking correction, 
$\delta_{C2}$, that used, on the one hand, Saxon-Woods eigenfunctions and, on the
other hand, Hartree-Fock eigenfunctions.  The former method, which was the one used by us
\cite{To77,TH02,TH08}, gave consistently larger corrections than the latter method, which
was used by Ormand and Brown \cite{OB85,OB89,OB95}.  This was deemed to be a systematic
effect caused by the different shapes of the Saxon-Woods and Hartree-Fock mean-field
potentials.  In this section, we re-address this issue for two reasons:  First, the
Hartree-Fock calculations of Ormand and Brown are now 14-24 years old and are not available
for all the nuclei under study here; and second, our most recent Saxon-Woods calculations
\cite{TH08} were performed in larger shell-model spaces not matched by the Hartree-Fock
calculations.  For a valid comparison both sets of calculations should be done in
identical shell-model spaces.

The isospin-symmetry-breaking correction, $\delta_{C2}$, for a superallowed $\beta$ transition
between $0^+$, $T$=1 analog states in nuclei with $A$ nucleons is computed from the formula
\cite{TH08}
\be
\delta_{C2} \simeq \sum_{\pi^<,\, \alpha} S_{\alpha}^< \Omega_{\alpha}^<
- \frac{1}{2}
\sum_{\pi^>,\, \alpha} S_{\alpha}^> \Omega_{\alpha}^> .
\label{dc2}
\ee
Here $S_{\alpha}$ is the spectroscopic factor for the pickup of a 
nucleon in quantum state $\alpha$ from an $A$-particle state
of spin $0^+$ and isospin $1$, to an $(A-1)$-particle state of
spin $\alpha$ and isospin $T_{\pi}$.  There is a complete-set
sum over all the $(A-1)$-particle states (called parent states, 
and denoted $\pi$) in Eq.~(\ref{dc2}).  The sum is divided into
two parts:  the first is over states with isospin $T_{\pi} = 1/2$ and
is denoted by $\pi^<$; the second is over states with $T_{\pi} = 3/2$,
denoted $\pi^>$.  Further, the $\Omega_{\alpha}^{\pi}$ are
radial-mismatch factors, which depend on the difference between
the radial wave function of a proton bound in the decaying nucleus,
$u_{p,\alpha}^{\pi}(r)$,
and that of a neutron bound in the daughter nucleus,
$u_{n,\alpha}^{\pi}(r)$.  Specifically, the radial-mismatch factors
are given by:
\bea
\Omega_{\alpha}^{\pi} & = & \int_0^{\infty}
u_{n,\alpha}^{\pi}(r)
\left [
u_{n,\alpha}^{\pi}(r)
-
u_{p,\alpha}^{\pi}(r)
\right ] ~ dr
\nonumber \\
& = & 1 - \int_0^{\infty}
u_{n,\alpha}^{\pi}(r)
u_{p,\alpha}^{\pi}(r) ~ dr .
\label{radover}
\eea
The radial functions are normalized to $\int | u(r) |^2 ~ dr = 1$.
They are labelled by the parent state $\pi$ because their asymptotic
forms are matched to their separation energies, which in turn depend
on the parent state.  For example, if the parent state is the 
ground state of the $(A-1)$-system, then the proton separation energy
would be $S_p$ and the neutron separation energy $S_n$, two
quantities given in terms of atomic masses and found in any
atomic mass table.  If, however, $\pi$ represents an excited
state of the $(A-1)$-system, then the proton and neutron
separation energies would be $S_p + E_x$ and $S_n + E_x$
respectively, where $E_x$ is the excitation energy of that
parent state.

To compute $\delta_{C2}$ from Eq.~(\ref{dc2}) one needs a set of
spectroscopic factors, $S_{\alpha}$, and a set of radial-mismatch
integrals, $\Omega_{\alpha}$.  The difference between Saxon-Woods and
Hartree-Fock calculations lies in the method used to evaluate the radial-overlap
integrals.  In one case, the radial functions $u(r)$ are taken to be eigenfunctions
of a Saxon-Woods potential; in the other case they are eigenfunctions
of a Hartree-Fock mean-field potential.  However, for a valid comparison
between them the $\delta_{C2}$ must be calculated for both with the 
same spectroscopic factors, $S_{\alpha}$.  This was not done in the 
past, it being assumed that the model spaces were
sufficiently comparable that this would not lead to any serious
error.  However, our most recent Saxon-Woods calculation \cite{TH08} have
significantly increased the model spaces over the ones used before, so
now it is essential that a new Hartree-Fock calculation be undertaken.  We report
such a calculation here.

To be clear about the procedure, we illustrate it for the specific case of
the decay of $^{34}$Cl to $^{34}$S.  The decaying nucleus has
$Z+1=17$ protons; the daughter nucleus has $Z$=16 protons.  In the
Saxon-Woods approach, the proton radial wave functions are taken to be
eigenfunctions of a potential defined for a nucleus of mass $A$
and charge $Z+1$ as follows:
\be
V(r) = - V_0 f(r) - V_s g(r) {\bf l}. \mbox{\boldmath$\sigma$}
+ V_C(r) ,
\label{VSW}
\ee
where
\bea
f(r) & = & \left \{ 1 + \exp \left ( (r-R)/a \right ) \right \}^{-1} ,  
\nonumber  \\
g(r) & = & \left ( \frac{\hbar}{m_{\pi} c} \right )^2 
\frac{1}{a_s r} \exp \left ( \frac{r - R_s}{a_s} \right )
\nonumber \\
& & ~~~~~ \times 
\left \{ 1 + \exp \left ( \frac{r - R_s}{a_s} \right )
\right \}^{-2} ,
\nonumber  \\
V_C(r) & = & Z e^2 / r , ~~~~ {\rm for}~~ r \geq R_c
\nonumber  \\
 & = & \frac{Z e^2}{2 R_c} \left ( 3 - \frac{r^2}{R_c^2} \right )
 , ~~~~ {\rm for}~~ r < R_c .
\label{Pot}
\eea
Here, $R = r_0 (A - 1)^{1/3}$ and
$R_s = r_s (A - 1)^{1/3}$.  
The three terms in Eq.~(\ref{VSW}) are the
central, spin-orbit and Coulomb terms respectively.  
In our calculations \cite{TH08} most of the parameters 
were fixed at standard values, with the well depth $V_0$
being adjusted case by case so that the binding energy of the
eigenfunction being computed matched the separation energy to
the corresponding parent state -- in $^{33}$S, for our example.  Likewise the
neutron radial functions were taken to be eigenfunctions of a similar
potential but with the Coulomb term omitted.

Although the Hartree-Fock procedure is comparable, there is one issue unique to this approach
which requires particular attention.  For our illustrative example, a Hartree-Fock
calculation might first be mounted for $^{34}$Cl, which would yield a mean field
with central, spin-orbit and Coulomb terms.  The required proton radial
functions would then be taken as eigenfunctions of this mean field with
the strength of the central term readjusted case by case so that
the computed binding energy matched the appropriate separation energy. 
A second Hartree-Fock calculation might then be mounted for $^{34}$S,
from which the neutron radial functions would be similarly determined in the
mean field, but without the Coulomb term.  However, under these circumstances,
if the Coulomb terms in the Hartree-Fock mean-field potential were to be compared with
those in the Saxon-Woods potential, a very significant difference would emerge.
In the Hartree-Fock case, the Coulomb term is
\be
V_C({\bf r}) = \int d^3{\bf r}^{\prime}~ \frac{e^2}{|{\bf r} -
{\bf r}^{\prime}|} \rho_p({\bf r}^{\prime}) -
\frac{3 e^2}{2} \left [ \frac{3}{\pi} \rho_p({\bf r}) \right ]^{1/3} ,
\label{coulHF}
\ee 
which depends on the proton density (of $^{34}$Cl in our example) that is
generated as part of the Hartree-Fock procedure.  The two terms in Eq.~(\ref{coulHF}) are
called the direct and exchange terms respectively.  If we take the asymptotic limit of the
direct term for large $r$, we obtain
\be 
V_C^{\rm dir}({\bf r}) =
\int d^3{\bf r}^{\prime} \frac{e^2}{|{\bf r} - {\bf r}^{\prime}|}
\rho_p({\bf r}^{\prime}) \stackrel{r \rightarrow \infty}{\longrightarrow}
\frac{e^2}{r} \int d^3{\bf r}^{\prime} \rho_p({\bf r}^{\prime})
= \frac{(Z+1)e^2}{r}.
\label{Vcdirect}
\ee
Since the Hartree-Fock proton density is normalized to $(Z+1)$ protons in $^{34}$Cl,
the asymptotic form of the Coulomb potential tends to $(Z+1)e^2/r$.  However, this
disagrees with the equivalent Saxon-Woods calculation, which has the form $Ze^2/r$
(see Eq.~\ref{Pot}).

This discrepancy is important and constitutes, in our opinion, a serious flaw
in this Hartree-Fock calculation of the radial-mismatch factor.  Since a
proton removed from a nucleus of charge $Z+1$ leaves behind $Z$ protons, its
asymptotic interaction is with charge $Z$ -- as described by the Saxon-Woods potential
-- and not with charge $Z+1$.  This deficiency in Hartree-Fock would be cured in
principle by the Coulomb exchange term.  However, in Skyrme-Hartree-Fock calculations
it is not possible to compute the exchange term exactly without sacrificing the
simplicities that come with use of zero-range Skyrme interactions.  The exchange term
appearing in Eq.~(\ref{coulHF}) is a commonly used local approximation, which 
might well be appropriate for the nuclear interior and for the computation of bulk
properties such as binding energies and radii, but it certainly does not do the job
asymptotically, which is the region of greatest importance to our calculations.  

To circumvent this difficulty, we have chosen to alter the Hartree-Fock protocol.  Instead
of mounting two Hartree-Fock calculations -- for $^{34}$Cl and $^{34}$S -- as just described,
we mount a {\it single} calculation for the nucleus with $(A-1)$ nucleons and $Z$ protons -- 
$^{33}$S in our example.  We then use the proton mean field from this calculation to
generate the proton eigenfunctions, $u_{p,\alpha}^{\pi}(r)$; and the neutron mean field from
the same calculation to generate the neutron eigenfunctions $u_{n,\alpha}^{\pi}(r)$.
In this procedure, the Coulomb interaction automatically has the correct asymptotic form. 
It is also fully consistent with the Saxon-Woods potential parameterisation, Eq.~(\ref{Pot}),
which considers the nucleus of mass $(A-1)$ as the core to which the last particle is bound,
since the radius of the potential is parameterised as $r_0 (A-1)^{1/3}$ rather than
$r_0 A^{1/3}$.  Calculations of $\delta_{C2}$ with this new Hartree-Fock protocol will be
presented in the next section.  It can be noted here, however, that these results are larger
than those obtained with the conventional protocol by between 10\% to 40\% depending on
the Skyrme interaction used and the nucleus under study.  This change of protocol goes a
long way in reducing the systematic error between Saxon-Woods and Hartree-Fock calculations.

\begin{table}[t]
\begin{center}
\caption{Adopted isospin-symmetry-breaking corrections, $\delta_{C2}$ in percent units, and
their assigned uncertainties obtained from Hartree-Fock calculations. Also listed are earlier
results obtained with Saxon-Woods (SW) eigenfunctions, as published in \protect\cite{TH08}. 
\label{t:dc2adopt}}
\vskip 1mm
\begin{ruledtabular}
\begin{tabular}{rcc}
& & \\[-3mm]
\multicolumn{1}{r}{Nucleus} &
\multicolumn{1}{c}{HF} &
\multicolumn{1}{c}{SW} \\[1mm]
\hline
& & \\[-3mm]
\mbox{\boldmath $T_z = -1:$} & & \\
$^{10}$C  & 0.215(35) & 0.165(15)  \\
$^{14}$O  & 0.255(30) & 0.275(15)  \\
$^{18}$Ne & 0.205(55) & 0.410(25)  \\
$^{22}$Mg & 0.250(55) & 0.370(20)  \\
$^{26}$Si & 0.335(55) & 0.405(25)  \\
$^{30}$S  & 0.540(55) & 0.700(20)  \\
$^{34}$Ar & 0.510(60) & 0.635(55)  \\
$^{38}$Ca & 0.600(60) & 0.745(70)  \\
$^{42}$Ti & 0.535(60) & 0.835(75)  \\[3mm]
\mbox{\boldmath $T_z = 0:$} & & \\
$^{26}$Al & 0.410(50) & 0.280(15)  \\
$^{34}$Cl & 0.595(55) & 0.550(45)  \\
$^{38}$K  & 0.640(60) & 0.550(55)  \\
$^{42}$Sc & 0.620(55) & 0.645(55)  \\
$^{46}$V  & 0.525(55) & 0.545(55)  \\
$^{50}$Mn & 0.575(55) & 0.610(50)  \\
$^{54}$Co & 0.635(55) & 0.720(60)  \\     
$^{62}$Ga & 0.93(16) & 1.20(20)   \\     
$^{66}$As & 1.10(35) & 1.35(40)   \\     
$^{70}$Br & 1.14(25) & 1.25(25)   \\     
$^{74}$Rb & 1.29(16) & 1.50(30)   \\[3mm]
\end{tabular}
\end{ruledtabular}
\end{center}
\end{table}
\subsection{New Hartree-Fock calculations for $\delta_{C2}$}
\label{ss:newHF}

Here we present our new Hartree-Fock calculations for the isospin-symmetry-breaking
correction, $\delta_{C2}$, for the 20 cases of superallowed Fermi beta decay
considered in our survey: they range from $^{10}$C to $^{74}$Rb.  Our procedure was 
that discussed at the end of the last section and involved obtaining the mean field
from a Hartree-Fock calculation in the $(A-1)$-system.  The proton and neutron radial
functions were obtained as eigenfunctions of this mean-field potential, whose overall
strength was scaled on a case-by-case basis to ensure the eigenfunction's asymptotic
solution matched the required separation energy.  These eigenfunctions were used to
compute the radial-mismatch factors, $\Omega_{\alpha}^{\pi}$ of Eq.~(\ref{radover}).
For the spectroscopic factors needed in Eq.~(\ref{dc2}), we ran several shell-model
calculations with the model spaces and effective interactions used recently \cite{TH08}
in calculations with Saxon-Woods potentials.  We also considered three choices of the
Skyrme interaction: SGII \cite{vGS81}, SkM* \cite{Ba82} and Ska \cite{Ko76}.  The first
two of these interactions were the ones used by Ormand and Brown \cite{OB89,OB95} in
their computations of $\delta_{C2}$.  The third interaction, Ska, is of similar quality
and was one of the first to be fitted to the incompressibility of nuclear matter, a key
constraint used in all later Skyrme interactions.  More recent Skyrme interactions tend
to be used in conjunction with pairing forces in Hartree-Fock-Boglioubov calculations
\cite{Ch97,Ch98,Kl08}.  Since we have not included pairing forces in the present work, we
have not attempted to use any of these more recent interactions.

The results from our Hartree-Fock calculations for $\delta_{C2}$ are listed in column 2 of
Table~\ref{t:dc2adopt}.  For each transition, the central value is an average of the
results obtained with the three choices of Skyrme interactions.  To assign an uncertainty,
we have examined the spread in results obtained with the different Skyrme interactions and
with different shell-model effective interactions and model spaces.  We also list in the
third column of the table the values we adopted from our Saxon-Woods computations, which
originally appeared in ref.~\cite{TH08}.

\end{document}